\documentclass[preprint,12pt]{aastex}
\usepackage{eucal}
\usepackage{epsf}

\newcommand{\beq}{\begin{equation}}           % begin equation
\newcommand{\eeq}{\end{equation}}             % end equation
\newcommand{\bef}{\begin{figure}}           % begin figure
\newcommand{\enf}{\end{figure}}             % end figure

\newcommand{\plotthree}[3]{\centering \leavevmode
\epsfxsize=.32\columnwidth \epsfbox{#1} \hfill
\epsfxsize=.32\columnwidth \epsfbox{#2} \hfill
\epsfxsize=.32\columnwidth \epsfbox{#3}}

\newcommand{\plotfourb}[4]{\centering \leavevmode
\epsfxsize=.22\columnwidth \epsfbox{#1} \hfill
\epsfxsize=.22\columnwidth \epsfbox{#2} \hfill
\epsfxsize=.22\columnwidth \epsfbox{#3} \hfill
\epsfxsize=.22\columnwidth \epsfbox{#4}}

\newcommand{\ploteight}[8]{\plotfourb{#1}{#2}{#3}{#4}\\
\plotfourb{#5}{#6}{#7}{#8}}

\shorttitle{Measuring Limb Darkening by Microlensing}

\received{2000 August 21}
\begin{document}

\title{Measuring Stellar Limb Darkening by Gravitational Microlensing}

\author{David Heyrovsk\'y\altaffilmark{1,2}}

\altaffiltext{1}{Harvard-Smithsonian Center for Astrophysics, 60
Garden St., Cambridge, MA 02138; \mbox{dheyrovsky@cfa.harvard.edu}}

\altaffiltext{2}{Na \v{S}\v{t}\'ahlavce 6, 160 00 Praha 6, Czech
Republic}

\begin{abstract}

Observations of microlensing transit events can be used to measure
the limb darkening of the lensed star. We discuss the advantages
and drawbacks of several microlensing light curve inversion
methods. The method of choice in this work is inversion by means of
decomposition of the stellar surface brightness profile. We
construct an ideal basis by principal component analysis of
brightness profiles obtained from model atmosphere calculations.
Limb darkening approximations using such a basis are superior to
those using standard power-law limb darkening laws. We perform a
full analysis of simulated single-lens microlensing transit events
including a detailed error analysis of the method. In realistic
events with a low impact parameter the brightness profile of the
source can be recovered with a relative accuracy of $2\%$ from the
center of the source disk to $0.9$ of the disk radius. We show that
in the particular case of the observed MACHO Alert 95-30 event the
intrinsic complex variability of the lensed red giant hinders efforts
to recover its surface features.

\end{abstract}


\section{INTRODUCTION}

Our current knowledge of the intrinsic physical properties of
stars is largely based on stellar atmosphere models
(e.g., \citealt{gray92}). Ever more advanced models are computed for
a broad range of stellar types, taking into consideration a range of
physical and chemical processes expected to occur in the star's
atmosphere.

However, for the vast majority of observable non-variable stars
the predictions of these models can be confronted with a single
observable quantity - the integrated full-disk spectrum. It
follows that any two physically different atmosphere models which
predict similar integrated spectra cannot be observationally
distinguished. More specifically, any observationally ``verified"
model may have a center-to-limb variation of the spectrum (or limb
darkening, in terms of broad-band photometry) very different from
the actual star and thus be unphysical, yielding wrong physical
parameters of the star. This inherent problem can be overcome by
testing atmosphere models in cases or events when the limb
darkening of the observed star can be measured.

Until recently the only stars other than the Sun providing the
opportunity to directly study their surfaces were the components
of eclipsing binaries. Measuring their limb darkening profiles
from eclipse light curves proved to be no easy task, due to
the degeneracy with other fitted parameters \citep{grygar72}. Even
obtaining useful constraints on the value of the linear limb
darkening parameter requires high quality light curves
\citep{popper84, popper85}.

Extra-solar planets occulting their parent stars are analogous to
eclipsing binaries with small secondaries. High precision
observations of these scarce systems can also be used for measuring
the parent star limb darkening. This was recently demonstrated in
the case of HD 209458, the first observed extra-solar occulting
system \citep{jha00,deeg01,brown01}.

The surfaces of nearby giant stars can be resolved by interferometry.
Developments in stellar interferometry have led to an advance from
measuring stellar radii \citep{mozur91} to detecting the effect of
limb darkening \citep{quirr96} to limb darkening measurement for
$\alpha$ Ori \citep{burns97}\footnote{This nearby supergiant was
also recently imaged directly using the Hubble Space Telescope
\citep{uidug98}.}. \citet{hajian98} points out the main difficulty
of such measurements -- signatures of differences between limb
darkening profiles become detectable only beyond the first zero of
the visibility function. At these high spatial frequencies the
fringe visibilities are low, and thus difficult to measure with
sufficient precision. Moreover, as shown by \citet{jacob00},
simultaneous fitting of the stellar radius and limb darkening is
complicated by degeneracy. These and other studies indicate that
while the use of interferometry for constraining the limb darkening
of the source is promising, the accuracy will not reach the level
achievable for eclipsing binaries in the near future.

In this work we concentrate on another method -- measuring limb
darkening from observations of source-transit microlensing events.
Galactic gravitational microlensing occurs when a foreground
massive object such as a dim star (hereafter ``lens") approaches
the line of sight to a background star (``source"). Due to
gravitational deflection of light by the lens, the flux from the
source is temporarily amplified as the lens passes in the
foreground \citep{pacz96}. Currently operating microlensing surveys
have already detected over 500 such events toward the rich stellar
fields of the Galactic bulge and the Magellanic Clouds. In case the
lens directly transits the source star, the light curve of the
event will be affected by its limb darkening profile
\citep{witt95,v-g95,ls95}. Observations of one such transit event,
MACHO Alert 95-30, were reported by \citet{alc97d}. By the time of
writing, limb darkening parameters have been measured also in four
binary-lens caustic-crossing events: MACHO 97-BLG-28
\citep{albrow99a}, MACHO 98-SMC-1 \citep{afonso00} \footnote{This
particular event yielded the first and thus far unique case of limb
darkening measurement for an A star from a metal-poor population.},
MACHO 97-BLG-41 \citep{albrow00} and OGLE-1999-BUL-23
\citep{albrow01}. \footnote{Preliminary limb darkening measurement
was reported in a fifth event, EROS BLG-2000-5 \citep{an02}, during
the review of this paper.}

Microlensing thus presents a unique new possibility for studying
the surfaces of distant stars \citep{v-g95,s96}. As discussed in \S
3, several approaches have been taken by other authors to extract
the limb darkening profile from microlensing light curves, with
encouraging results. Here we present another inversion strategy
providing the most accurate and stable results so far. We achieve
this by decomposing the profile using a basis constructed by
principal component analysis of a set of actual model atmosphere
profiles.

In the following presentation, \S 2 contains the basic description
of single point-mass lens microlensing of a stellar source. In \S 3
we discuss several methods of inverting the microlensing light curve
to obtain the surface brightness profile of the source.
Specifically, in \S 3.3 we provide the basic equations for
inversion by brightness profile decomposition. Using model
atmosphere calculations and principal component analysis of the
model profiles, we determine an ideal basis for brightness
decomposition in \S 4. The main results are presented in \S 5, in
which we analyze simulated light curves, simultaneously recovering
the lensing parameters and the brightness profile of the source. We
perform a detailed error analysis and compute the inversion
accuracy for a source with unknown limb darkening. We compare the
results with those obtained by fitting the linear limb darkening
law, applied in all five published events (an additional
square-root coefficient was fitted in MACHO 97-BLG-28). The
analysis of the observed MACHO Alert 95-30 event light curves is
described in \S 6. Limb darkening measurement prospects are
discussed in \S 7, followed by a summary of the main results in \S
8. Two appendices are added, the first describing broadband
chromaticity of microlensing light curves, the second discussing
the model limb darkening profiles and their fits.

\section{GRAVITATIONAL MICROLENSING LIGHT CURVES}

Throughout this work we describe the source and lensing geometry
in terms of angular displacements in the plane of the sky, using
the angular radius of the lensed star as a (projected) distance
unit in this plane.\footnote{Using the angular Einstein radius as
a unit (traditional microlensing notation) is not convenient in
this work concentrating on the source star structure. The
conversions between our parameters and the parameters in the
standardized notation proposed by \cite{gould00} are
$l=u\theta_E/\theta_*$, $\epsilon=\theta_E/\theta_*$,
$p=u_0\theta_E/\theta_*$ and $t_*=t_E\theta_*/\theta_E$.} We
concentrate here on the case of a spotless source star, with a
symmetric surface brightness profile $B(r)$. Microlensing of a
spotted source is discussed elsewhere \citep{hs00a}.

A point-mass lens located at a projected distance $l$ from the center of
the disk of a background star increases the observed flux from this source
star to a value
\beq
F(l)=\int\limits_0\limits^1 B(r)\,\mathcal{A}(l,r)\,r\,dr\quad ,
\label{eq:main}
\eeq
where $r$ is the distance from the center of the star, and the surface
brightness is measured in appropriate units. The angle-integrated
amplification $\mathcal{A}(l,r)$ is defined by
\beq
\mathcal{A}(l,r)=\int\limits_0\limits^{2\pi} A_0
(\sqrt{l^2-2\,r\,l\cos{\phi}+r^2})\,d\phi\quad,
\label{eq:angint}
\eeq
where $\phi$ is the (source-centered) polar angle between a point on the
stellar disk at a distance $r$ and the lens at a distance $l$. The lensing
amplification factor $A_0(x)$ for a point source at a separation $x$ from
the lens is
\beq
A_0(x)=\frac{x^2+2\,\epsilon^2}{x\sqrt{x^2+4\,\epsilon^2}}\quad ,
\eeq
here $\epsilon$ is the angular Einstein radius of the lens \citep{pacz96}
in units of lensed star radius. The integral in
equation~(\ref{eq:angint}) can be directly solved in terms of elliptic
integrals as follows:
\begin{eqnarray}
\mathcal{A}(l,r)=\frac{4}{(l+r)\sqrt{(l-r)^2+4\,\epsilon^2}}\;
&\Bigg[&2\,\epsilon^2\;K\Bigg(\,\frac{4\,\epsilon}{l+r}
\sqrt{\frac{l\,r}{(l-r)^2+4\,\epsilon^2}}\,\Bigg)
+{}\nonumber\\\nopagebreak
{}&+&(l-r)^2\;\Pi\Bigg(\,\frac{4\,l\,r}{(l+r)^2}\,,\,\frac{4\,\epsilon}{l+r}
\sqrt{\frac{l\,r}{(l-r)^2+4\,\epsilon^2}}\,\Bigg)\Bigg] .
\label{eq:aiampl}
\end{eqnarray}
We use standard notation for the complete elliptic integrals of the first
($K$) and third ($\Pi$) kind. In the latter we follow the sign convention
used by \cite{byrdfried71}\footnote{ \cite{press92} use an opposite sign
for the first parameter.},\\
\mbox{$\Pi(\alpha^2,k)=\int_0^{\pi/2}(1-\alpha^2\sin^2{\theta})^{-1}
(1-k^2\sin^2{\theta})^{-1/2}\,d\theta\,$.}

Notice that $\mathcal{A}(l,r)$ is symmetric in its arguments
$l,r$. Also note that $\mathcal{A}(0,r)= 2\pi A_0(r)$, as well as
$\mathcal{A}(l,0)= 2\pi A_0(l)$. When the lens is far from the
source ($l\gg 1, \epsilon$), the angle-integrated amplification
drops to $2\pi$. In the limit of a strong lens ($\epsilon\gg 1$),
the angle-integrated amplification at the light curve peak
($\epsilon\gg l$) in equation~(\ref{eq:main}) can be approximated
by
\beq
\mathcal{A}(l,r)\simeq \frac{4\,\epsilon}{l+r}
K\Big(\,\frac{2\sqrt{l\,r}}{l+r}\,\Big)\quad ,
\label{eq:limaiampl}
\eeq
an expression linearly dependent on the
Einstein radius $\epsilon$. In this work we use the more general
expression in equation~(\ref{eq:aiampl}). Figure~\ref{fig:aiampl}
contains sample plots of the angle-integrated amplification as a
function of radial distance $r$ for Einstein radii $\epsilon=$ 1
and 10 (the latter scaled down by a factor of 10 for comparison),
and several lens positions. At the position of the lens ($r=l$)
there is a logarithmic divergence, so that for $r=l+\delta$
\beq
\mathcal{A}(l,l+\delta)\simeq
\frac{2\,\epsilon}{l} \Big(1-\frac{\delta}{2\,l}\Big)
\ln{\frac{8\,\epsilon\,l}{|\delta|\sqrt{l^2+\epsilon^2}}}
+4\arctan{\frac{l}{\epsilon}} +
\frac{\epsilon\,(2\,l^2+\epsilon^2)}{l^2(l^2+\epsilon^2)}\,\delta
+\mathcal{O}(\delta^2\ln{|\delta|}) \quad ,
\label{eq:divergence}
\eeq
where $|\delta|\ll 1$. Due to this divergence at the lens
position (for $l\leq 1$) in the integrand of
equation~(\ref{eq:main}), the lens efficiently scans the surface
brightness profile of the source star as it moves across the
stellar disk. The general microlensing effect on the spectrum of
model stellar atmospheres was demonstrated by
\citet[hereafter HSL]{hsl00}. We provide a brief discussion of
broadband light curve shapes in Appendix A of this work.

\section{LIGHT CURVE INVERSION METHODS}

Photometric observations of microlensing events provide us with
sets of observed fluxes $F_i$ measured at different times $t_i$
with estimated errors $\sigma_i$ ($i=1,\ldots n$). Primary
analysis of a transit event yields the main (geometric) lensing
parameters: the Einstein radius $\epsilon$, impact parameter $p$,
time of closest approach $t_0$ and angular source radius crossing
time $t_*$. The latter three parameters help us mainly to convert
the observation times $t_i$ to lens positions $l_i\,$, as shown
further in equation~(\ref{eq:lenspos}). In the two following
sections (through \S 4), we assume zero blended flux (contribution
of unresolved stars) and consider the geometric parameters to be
fixed at predetermined values, in order to facilitate the
inversion analysis. In \S 5 and \S 6 we proceed to the full event
analysis including the effect of blending, and recover the lensing
and limb darkening parameters simultaneously.

The aim of this work is to extract the brightness profile $B(r)$
from the observed fluxes $F_i$. The mathematical problem is to
invert equation~(\ref{eq:main}), classified as a Fredholm integral
equation of the first kind. This problem has no universal method
of solution. The task is more difficult in this particular case,
because of the required high level of accuracy of the result. The
aim is not just to detect limb darkening, but to actually measure
it. Based on solar observations and model atmosphere calculations,
broadband limb darkening profiles in the optical range vary only
to a fairly limited degree (see Appendix B), hence the required
precision.

There is a number of different numerical approaches to inverting
equation~(\ref{eq:main}) that can be tested. \cite{crbr86} provide
a good overview of strategies for solving astrophysical inverse
problems. In the following two subsections we discuss two methods
applied previously by other authors to the microlensing case. The
third subsection introduces the method used in this work.

\subsection{Inversion by Discretization}

A straightforward method is to invert equation~(\ref{eq:main}) by
discretization, approximating the integral by a sum. For a light
curve with $n$ points we can select $n_D\leq n$ radial points
$r_j$ ($j=1,\ldots n_D$) on the source disk and write \beq
F(l_i)=\int\limits_0\limits^1
B(r)\,\mathcal{A}(l_i,r)\,r\,dr=\sum\limits_{j=1}\limits^{n_D}
\mathbb{D}_{ij} \breve{B}(r_j) \label{eq:discret} \eeq for each
lens position $l_i$ ($i=1,\ldots n$). Here the matrix element
$\mathbb{D}_{ij}$ is some approximation of
$\int\mathcal{A}(l_i,r)\,r\,dr$ over the vicinity of point $r_j$.

In order to obtain brightness point estimates $\breve{B}(r_j)$ providing a
good fit to the light curve data, we now minimize
\beq
\chi^2=\sum\limits_{i=1}\limits^{n} \sigma_i^{-2} [F_i - F(l_i)]^2\quad .
\label{eq:chi2}
\eeq
Rescaling the matrix to $\widetilde{\mathbb{D}}_{ij}\equiv\mathbb{D}_{ij}
/\sigma_i$ and the measured fluxes to $\tilde{F}_i\equiv F_i/\sigma_i$, we
can write the $\chi^2$-minimizing solution explicitly:
\beq
\breve{B}(r_j)=\sum\limits_{i=1}\limits^{n}\sum\limits_{k=1}\limits^{n_D}
\big(\widetilde{\mathbb{D}}^{\,T}\widetilde{\mathbb{D}}\big)^{-1}_{jk}\,
\widetilde{\mathbb{D}}_{ik} \tilde{F}_i \quad ,
\label{eq:discrsol}
\eeq
here $\widetilde{\mathbb{D}}^{\,T}$ denotes the transpose of matrix
$\widetilde{\mathbb{D}}$. The errors due to light curve noise for each of
the obtained brightness points are given by the standard deviations
\beq
\sigma_{\breve{B}_j} =
\sqrt{\big(\widetilde{\mathbb{D}}^{\,T}\widetilde{\mathbb{D}}\big)^{-1}_{jj}}
\quad .
\eeq
In comparison with the other methods discussed subsequently, this
direct approach has the advantage of not making explicit
assumptions about the shape of the brightness profile.

\citet{gaugou99} presented an error analysis of this
method\footnote{ Their paper covers the case of binary lensing as
well, in the linear caustic crossing regime. Here we discuss their
single lens results.}, demonstrating that the surface brightness
profile can be recovered only for the outer radial part of the
stellar disk directly transited by the lens. The profile of the
inner part, at radii smaller than the impact parameter of the
lensing event, cannot be accurately reconstructed.

The general failure of microlensing light curve inversion in this
region can be explained as follows. For any light curve point
$l_i>0$, the matrix element $\mathbb{D}_{ij} \rightarrow0$ for
inner disk points $r_j\rightarrow0$ by definition. Hence, in the
case of a non-zero impact parameter the associated value
$\breve{B}(r_j)$ has negligible weight in all the sums given by
equation~(\ref{eq:discret}), and thus is weakly constrained by the
data. Alternatively, this failure can be understood by noting the
low microlensing sensitivity to stellar spots located off the
projected lens trajectory (e.g., a circular spot at the disk
center; see \citealt{hs00a}).

Returning specifically to inversion by discretization, some of its
intrinsic problems are directly apparent from
equation~(\ref{eq:discret}). For transit points ($l_i<1$) the
integrand is divergent, as demonstrated by
equation~(\ref{eq:divergence}). Obtaining an adequate
approximation of such an integral by a sum requires a fairly high
number of terms, i.e., an even higher number of transit light
curve points. Even in the case of negligible light curve noise,
simulations show that discretization with a limited number of
points introduces errors that make the method unstable and the
inversion unreliable.

As a further point, straightforward discretization assuming
constant surface brightness within annuli of the source disk can
introduce additional errors. Unless there is a sufficient number
of radial points defining the annuli, this assumption cannot be
made particularly near the limb of the source, where the
brightness drops rapidly.

The stability of this method can be improved by introducing additional
constraints on the solution points $\breve{B}(r_j)$. For example, one
might require positiveness of the solution, its monotonic decrease or
concavity. Typically these constrain the solution points by a set of
linear inequalities. The solution is then found numerically by minimizing
$\chi^2$ given by equation~(\ref{eq:chi2}), while satisfying the imposed
inequalities. \citet{bogcher96} took this approach in the case of
non-transit events\footnote{ This choice is not ideal, as non-transit
events contain little information about the surface brightness
distribution of the source (see HSL and Appendix A).}, seeking a
non-negative concave decreasing solution. Unfortunately, they made no
estimate of the errors or stability of their results. In a following paper
\citep{bogcher99}, the authors presented their inversion of the MACHO
Alert 95-30 light curves, by seeking a non-negative decreasing solution.
No stability estimates were made in this work either, moreover the data
manipulation is questionable (neglecting observational errors, converting
fluxes to amplifications prior to inversion etc.).

Even the constrained method has its implicit drawbacks. Either the
unconstrained solution given by equation~(\ref{eq:discrsol}) automatically
satisfies the desired constraints, or the constrained solution has to
satisfy at least one equality instead of the required inequality. This is
a direct consequence of the minimized function $\chi^2$ being quadratic
and positive definite in its parameters $\breve{B}(r_j)$. Such a function
has only one global minimum, given by the unconstrained
solution~(\ref{eq:discrsol}), and no other local minima. If this minimum
does not lie within the constrained region of parameter space, the
constrained solution lies on the boundary of the region. A requirement of
positiveness therefore forces the constrained solution to have at least
one zero point, a requirement of monotonic decrease forces a constant
section of the solution, concavity forces a linear section, etc. A
combination of constraints forces an equality in at least one constraint.
Even if the method provides reasonable-looking solutions, there is no
guarantee these approximate the actual brightness profile of the source.
Without an error analysis based on simulations the results cannot be
trusted.

\subsection{Backus-Gilbert Method}

The Backus-Gilbert method (e.g., \citealt{crbr86}) takes quite a different
approach to inverting equation~(\ref{eq:main}). It sets out to determine
an approximation of the brightness at some point $r_0$ as a linear
combination of all the observed fluxes,
\beq
\bar{B}(r_0)=\sum\limits_{i=1}\limits^{n} \mu_i(r_0)\,F_i \quad ,
\label{eq:bgappr}
\eeq
where $\mu_i(r_0)$ are suitable coefficients. Substituting the theoretical
expectations $F(l_i)$ for the fluxes from equation~(\ref{eq:main}), we
find
\beq
\bar{B}(r_0)\simeq\int\limits_0\limits^1 S(r,r_0)\,B(r)\,dr \quad ,
\label{eq:bgconv}
\eeq
where the introduced spread function $S(r,r_0)$ is a linear combination of
the integral kernels corresponding to all lens positions in
equation~(\ref{eq:main}), using the same coefficients,
\beq
S(r,r_0)\equiv\sum\limits_{i=1}\limits^{n} \mu_i(r_0)\,
\mathcal{A}(l_i,r)\,r \quad .
\label{eq:bgspread}
\eeq
The spread function should be normalized so that a uniform brightness
profile is recovered identically by equation~(\ref{eq:bgconv}), namely
$\int_0^1 S(r,r_0)\,dr=1$. The same equation suggests it is desirable to
select the coefficients $\mu_i(r_0)$ so that the spread function
approaches the delta function $\delta(r-r_0)$. This narrowness (high
resolution) of the spread function is sought by requiring for example
\beq
\int\limits_0\limits^1 (r-r_0)^2\,S^2(r,r_0)\,dr = Min.
\label{eq:bgmin}
\eeq
The final constraint imposed on the coefficients limits the error
magnification,
\beq
\sigma^2_{\bar{B}(r_0)}=\sum\limits_{i=1}\limits^{n} \mu^2_i(r_0)\,
\sigma^2_i \leq M\,\sum\limits_{i=1}\limits^{n} \sigma^2_i \quad ,
\label{eq:bglim}
\eeq
the constant $M$ is a measure of the statistical error tolerance. The set
of coefficients $\mu_i(r_0)$ can now be easily found analytically by
minimizing equation~(\ref{eq:bgmin}) using the definition from
equation~(\ref{eq:bgspread}) together with the normalization condition and
the error magnification constraint. We then compute the brightness point
$\bar{B}(r_0)$ from equation~(\ref{eq:bgappr}) and its error due to light
curve noise from equation~(\ref{eq:bglim}).

The method tends to be laborious -- the procedure has to be repeated for
each desired point $\bar{B}(r_0)$ individually. Another drawback lies in
having the free parameter $M$. Decreasing its value on one hand increases
the stability of the result towards light curve noise, but on the other
hand decreases the resolution (yields a broader spread function). As a
result, some user-defined optimization is necessary. It should be noted
that there is no explicit requirement that the obtained solution provide a
good fit to the light curve data. The result may be stable towards light
curve noise and yet be incorrect due to systematic effects. This should be
checked independently.

Preliminary reports on testing the Backus-Gilbert method were presented by
\citet{hendry98} with a recent update by \citet{graco00}. In both cases
their presented sample result is for a lens with an Einstein radius
$\epsilon=1$ and a non-transit impact parameter $p=1\,$. It should be
noted that for these parameters the surface brightness signature on the
light curve is very weak (see HSL and Appendix A), in fact lower than the
2$\%$ noise added to the simulated light curve in
\citet{hendry98}\footnote{ \citet{graco00} show that the results do not
change qualitatively if the light curve noise is varied from $0.2\%$ to
$10\%$, which suggests the result is an artifact of the method rather than
a measurement of the source profile.}. Using a larger Einstein radius
(which is also more realistic) should provide better results. At radii
$0.6\lesssim r \lesssim 0.9$ the recovered brightness approaches the
original profile, though there is a systematic discrepancy in the
residuals. Between $0.5$ and $0.6$ the obtained profile peaks and starts
decreasing (unphysically) towards smaller radii, reaching a $\sim -18\%$
deviation at $r=0.05$ (even worse, $\sim -25\%$ in \citealt{graco00}). As
a consequence, the recovered profile has a lower integrated flux, and thus
will not provide a good fit to the observed light curve. The authors state
that the deviation for the region $r<0.6$ does not improve with reduced
light curve noise. The inversion turns unacceptably noisy for 10 or less
light curve points according to the authors. It is apparent that in this
case the actual inversion error is determined by systematic effects rather
than by statistical errors. The accuracy is too low for the results to be
of much use for measuring the limb darkening of the source.

The Backus-Gilbert technique seems to be poorly suited for microlensing
light curve inversion in general. In the case of non-transit events, all
the kernels on the right-hand side of equation~(\ref{eq:bgspread}) are
very similar functions, increasing monotonically from the source center
towards the limb. It is difficult to construct a spread function narrowly
peaked at any selected point on the disk by a linear combination of such
terms. At any rate, this would require at least some of the coefficients
$\mu_i(r_0)$ to be negative. However, subtracting fluxes in
equation~(\ref{eq:bgappr}) only increases the effects of noise. The
problem in recovering the brightness profile of the inner disk is also
apparent: at $r=0$ all the kernels are identically equal to zero. Getting
a spread function peaked in this region would require high values of the
coefficients, which would increase its amplitude at the limb even more.

In the case of transit lens positions, the corresponding kernels in
equation~(\ref{eq:bgspread}) are divergent at these points. Getting a
spread function at least effectively narrow-peaked at a particular point
by a weighted combination of functions divergent at different points seems
to be difficult. It remains to be seen whether and how the Backus-Gilbert
technique could be modified to be useful for microlensing light curve
inversion.

\subsection{Limb Darkening Decomposition}

We can avoid a number of problems encountered by the previous methods by
decomposing  the brightness profile into a linear combination of $n_C\leq
n$ basis functions $f_j(r)$,
\beq
B(r)\simeq\hat{B}(r)=\sum\limits_{j=1}\limits^{n_C} \alpha_j f_j(r)
\quad .
\label{eq:decomp}
\eeq
We can now rewrite equation~(\ref{eq:main}),
\beq
F(l_i)=\int\limits_0\limits^1 B(r)\,\mathcal{A}(l_i,r)\,r\,dr\simeq
\sum\limits_{j=1}\limits^{n_C}\mathbb{C}_{ij} \alpha_j \quad ,
\label{eq:decompflux}
\eeq
where the matrix elements $\mathbb{C}_{ij}$ are defined by
\beq
\mathbb{C}_{ij}\equiv \int\limits_0\limits^1\mathcal{A}(l_i,r)\,f_j(r)
\,r\,dr \quad ,
\label{eq:decmatrix}
\eeq
and can be therefore computed to arbitrary precision. We now use
equation~(\ref{eq:decompflux}) to minimize $\chi^2$ given by
equation~(\ref{eq:chi2}), obtaining the best-fit coefficients
\beq
\alpha_j=\sum\limits_{i=1}\limits^{n}\sum\limits_{k=1}\limits^{n_C}
\big(\widetilde{\mathbb{C}}^{\,T}\widetilde{\mathbb{C}}\big)^{-1}_{jk}
\,\widetilde{\mathbb{C}}_{ik} \tilde{F}_i \quad .
\label{eq:coefs}
\eeq
As previously, we rescaled the matrix elements to
$\widetilde{\mathbb{C}}_{ij} \equiv\mathbb{C}_{ij}/\sigma_i$. The standard
deviations of the obtained coefficients due to light curve noise are
\beq
\sigma_{\alpha_j} =
\sqrt{\big(\widetilde{\mathbb{C}}^{\,T}\widetilde{\mathbb{C}}\big)^{-1}_{jj}}
\quad .
\eeq

This method is mathematically closely related to inversion by
discretization. However, unlike in the case of discretization, the
involved integrals can be computed to arbitrary precision and
therefore do not contribute to the inversion error. The quality of
this method is given by the accuracy of the decomposition in
equation~(\ref{eq:decomp}), which is the only introduced
approximation. In principle, with a sufficiently high number of
decomposition terms the range of brightness profile shapes is
restricted only negligibly. However, with a limited number of terms
this method is clearly model-dependent, unlike discretization. In the
following section we proceed to construct a basis $f_j(r)$ suitable
for inverting broadband light curves.

\section{SELECTION OF DECOMPOSITION BASIS}

To test the decomposition method, we generate simulated
microlensing light curves using sample brightness profiles obtained
directly from model atmosphere computations (see HSL and
\citealt{hl97} for more details). We explore a principally
four-dimensional parameter space, setting the Einstein radius
$\epsilon$, impact parameter $p$ (position of the closest light
curve point here), number of light curve points $n$ (in most cases
only transit points are used), and relative noise level $\eta$. We
add noise to the simulated light curve by setting
\beq
F_i=F(l_i)\,[\,1+\eta_i\,] \quad ,
\eeq
where $\eta_i$ ($i=1,\ldots n$) is a Gaussian-distributed random
number with zero mean and standard deviation $\eta$. Hence
$\eta\,F(l_i)$ represents the error $\sigma_i$.

For a given basis $f_j(r),\, j=1,\ldots n_C\,$ we evaluate the
quality of the inversion $(B/\hat{B}-1)$ and compare it with
results obtained for the standard limb darkening (hereafter SLD)
basis
\beq
g_1(r)=1\;,\quad
g_j(r)=-\big[\,1-\sqrt{1-r^2}\,\big]^{\,j-1}\quad \mbox{for}\;
j>1\;.
\label{eq:lld}
\eeq
In this basis two terms ($n_C=2$) correspond to the usual linear
limb darkening law, three to the quadratic law, one to uniform
brightness. Four out of the five published microlensing limb
darkening measurements mentioned in \S 1 used the linear law.
\citet{albrow99a} measured an additional term
($1-\sqrt[4]{1-r^2}$, square-root law) in MACHO 97-BLG-28.

In the following subsection we illustrate some general features
and problems of this method on the case of an analytical basis. In
\S 4.2 we derive an ideal basis by principal component analysis of
a set of brightness profiles obtained from realistic model
atmosphere calculations.

\subsection{Polynomial Basis}

We first try to set $f_j(r)=P_j(r)$, where $P_j(r)$ is a
polynomial satisfying the requirement $dP_j/dr(0)=0\,$, $P_1(r)$
is a constant, and $P_j(r)$ is of degree $j$ for $j>1$.\footnote{
The inversion result $\hat{B}(r)$ is independent of the explicit
form of the basis satisfying these criteria.} Any brightness
profile can be well approximated using several terms of such a
basis. In the following we briefly summarize the main results of
inversion tests.

The brightness profile close to the limb is generally poorly
constrained by light curve data. For practical purposes we
concentrate on the inversion accuracy for the interior region
$r\lesssim0.9$ of the stellar disk.

As one would expect from the microlensing chromaticity curve (see
Appendix A), the lower the impact parameter, the better the
inversion. While non-transit points with small errors can help
improve the inversion accuracy, transit points are necessary for
achieving an accuracy better than 10\%. The inversion quality is
independent of the Einstein radius for $\epsilon\gtrsim 3$, it
deteriorates slowly for $\epsilon<3$. This is in agreement with
the microlensing sensitivity to surface structure, which doesn't
increase much for $\epsilon\gtrsim 3$ while for smaller
values\footnote{ As noted in HSL, for practical purposes the lower
cutoff is $\epsilon\simeq 0.4$. Events with smaller Einstein radii
are too weak (maximum amplification $\lesssim 1.34$) to trigger an
alert in most current surveys.} it drops to zero, as illustrated
in Appendix A.

The inversion accuracy rapidly deteriorates when using more than a
couple of basis terms. Moreover, the number of terms necessary for
obtaining the best inversion is not correlated with the changing
$\chi^2/(degree\, of\, freedom)$. As one cannot directly judge the
quality of the inversion by the quality of the light curve fit, it
is necessary to predetermine an ideal number of terms to extract
from a light curve with given parameters. Generally, we find it is
possible to extract 4 polynomial terms for a noise level of 0.1\%,
3 terms for a currently achievable level of 0.5\%, and merely 2
terms (i.e., a parabola) for more common noise levels of 1\% and
2\%. In the SLD case, only 2 terms (i.e., standard linear limb
darkening) are reliably obtainable in general.

Overall, the SLD inversions tend to be superior to the tested
polynomial inversions. However, the quality of the SLD results
depends on the underlying brightness profile $B(r)$. The results
can be systematically worse in profiles poorly described by linear
limb darkening. The tested polynomial basis is not ideal either.
It describes a too general class of functions, so that it is not
possible to extract a sufficient number of terms to provide a
reliable and useful light curve inversion. From these tests we
conclude we need a basis significantly more constrained than
general polynomials, but capable of describing already in {\em two
terms} a broader range of limb darkening profiles than does
standard linear limb darkening.

\subsection{Basis Constructed by Principal Component Analysis}

For constructing an adequate basis we start out with $BV\!RI$
profiles obtained from eight computed red giant model atmospheres
(see Appendix B and Fig.~\ref{fig:profiles}), using filter
response functions provided by \citet{bessell90}. Our goal is to
find a basis giving the best gradual (single-term, two-term,
$\ldots$) approximation to these $n_M=32$ different brightness
profiles $B_i(r),\, i=1,\ldots n_M\,$. We can obtain the first
normalized component of the basis ($\int_0^1 f_1^2(x)\,dx = 1\,$)
by minimizing the total integrated square residual
\beq
\mathcal{R}(f)=\sum\limits_{i=1}\limits^{n_M}\,
\int\limits_0\limits^1\,\Bigg\{b_i(x)-\bigg[\int\limits_0\limits^1 b_i(y)
f(y) \,dy \bigg]\; f(x) \Bigg\} ^2 \; dx \quad ,
\eeq
where $b_i(r)$ are the model profiles, suitably normalized. The inner
integral is the coefficient approximating $b_i(x)$ by $f(x)$. The
minimization can be performed by the calculus of variations, leading to
an equation of the form
\beq
\int\limits_0\limits^1\,\sum\limits_{i=1}\limits^{n_M}b_i(x)b_i(y)\,f(y)
\,dy = \lambda_f\, f(x) \quad.
\label{eq:eigen}
\eeq
We can see that the sought function $f_1(r)$ is an eigenfunction of a
positive semidefinite integral operator with a symmetric kernel. It can
be shown not only that $f_1(r)$ corresponds to the highest eigenvalue
$\lambda_{f_1}\,$, but also that all the further components of the basis
can be obtained by extracting normalized eigenfunctions with successively
lower eigenvalues\footnote{ As a bonus, orthogonality of the
eigenfunctions, $\int_0^1 f_i(x) f_j(x)\,dx = \delta_{ij}\;$, is
guaranteed, because the operator in equation~(\ref{eq:eigen}) is
Hermitian.}. The total integrated square residual for a $k$-component
basis is
\beq
\mathcal{R}(f_1,\ldots f_k)=
\sum\limits_{i=1}\limits^{n_M}\,\int\limits_0\limits^1\,\Bigg\{
b_i(x)-\sum\limits_{j=1}\limits^k \bigg[\int\limits_0\limits^1 b_i(y)
f_j(y) \,dy \bigg]\; f_j(x) \Bigg\} ^2 \; dx \quad ,
\label{eq:pcares}
\eeq
its minimization leads to equation~(\ref{eq:eigen}) for each of the
components independently, thanks to their orthogonality. The eigenvalue
for the $j$-th component is equal to the decrement of the residual
$\mathcal{R}$ due to the addition of the $j$-th component to the basis,
\beq
\lambda_{f_j}=\sum\limits_{i=1}\limits^{n_M}\bigg[\int\limits_0\limits^1
b_i(x)\,f_j(x)\,dx \bigg]^2 \quad .
\eeq
By substituting the constructed eigenfunction basis into
equation~(\ref{eq:pcares}), we obtain the actual value of the total
residual,
\beq
\mathcal{R}(f_1,\ldots f_k)=\sum\limits_{i=1}\limits^{n_M}
\int\limits_0\limits^1 b_i^2(x)\,dx - \sum\limits_{j=1}\limits^k
\lambda_{f_j} \quad .
\eeq
From this expression we can see that the normalization of the brightness
profiles which gives each an equal weight in this analysis is
\beq
b_i(r)=\bigg[\int\limits_0\limits^1 B_i^2(y)\,dy\bigg]^{-1/2}\,B_i(r)
\quad ,
\eeq
and consequently $\mathcal{R}(f_1,\ldots f_k) = n_M - \sum_{j=1}^k
\lambda_{f_j}$. If there are $n_I$ linearly independent profiles
among $B_i(r)$, there are exactly $n_I$ positive eigenvalues - all
the rest are zero. Their sum is $n_M$, hence they leave a zero
residual. In most cases based on non-trivial model atmosphere
calculations (such as in this case), all $B_i(r)$ are linearly
independent and $n_I=n_M$.

The approach sketched above is a special case of principal
component analysis (e.g., \citealt{preis88,joll86}). The
eigenfunction problem involves solving the homogeneous Fredholm
integral equation~(\ref{eq:eigen}), which can be easily achieved
numerically by iteration. The first three basis functions obtained
by principal component analysis of our 32 model profiles are
presented in Figure~\ref{fig:basis}.

We demonstrate the superiority of the first two components of the
basis over linear limb darkening in Appendix B, for the model
profiles as well as for data from optical observations of the Sun.
This result suggests that although the basis was constructed from
brightness profiles of cool red giants, it can be used for a
broader class of sources. On the other hand, similar bases can be
constructed using sets of model atmospheres of different stellar
types. The main reasons for using red giants in this work are
their intrinsic brightness (observational advantage of achieving
better photometry) and large size (advantage of getting better
coverage of the transit). In addition, red giants are numerous in
the Galactic bulge fields observed by the microlensing survey
teams. Finally, the only well-documented single lens microlensing
transit event by the time of writing, MACHO Alert 95-30
\citep{alc97d}, involved a red giant source.

Initial inversion tests with the basis obtained by principal
component analysis (hereafter PCA) and fixed lensing parameters as
in the previous subsection show that the systematic errors plaguing
the SLD inversions are significantly reduced. The obtained results
are generally more accurate and less biased. However, it is
necessary to test PCA inversion in the full light curve analysis,
without knowing the lensing parameters a priori.

\section{FULL INVERSION OF SIMULATED LIGHT CURVES}

In order to demonstrate the reliability and quality of the described PCA
inversion, we test the method for a representative range of lensing
parameter combinations and source brightness profiles. We estimate the
statistical error by inverting a number of noisy light curves for each
parameter combination, and the systematic error from the results obtained
for different brightness profiles. For comparison, we perform the same
procedure using linear limb darkening (hereafter LLD).

In the following subsection we describe the implementation of the
simulations and the light curve inversion algorithm. In \S 5.2 we
discuss the accuracy of the recovered lensing parameters, and in \S
5.3 the obtained limb darkening of the source. Based on these
results, we suggest in \S 5.4 how to analyze observed light curves
in order to accurately recover the limb darkening as well as the
lensing parameters.

\subsection{Simulation and Inversion Algorithms}

We generate simulated light curves for combinations of five
lensing parameters: Einstein radius $\epsilon$, impact
parameter $p$, blend parameter $\beta$ (ratio of flux from
unresolved stars to baseline flux $F_*(\infty)$ of lensed
star), number of points per source diameter crossing time
$n_*$ (equal to transit point number for small impact parameters),
and light curve noise level $\eta$. The two parameters
defining the time scale of the simulated event can be kept fixed
at arbitrary values, for simplicity we set the source radius
crossing time $t_*=1$ and the time of closest approach $t_0=0$.
However, we do fit for both of these parameters in the following
light curve analysis. In order to get a reasonable fit for most of
the parameters we include non-transit points, setting the total
number of points to $n=2\,n_*$ and distributing them in equal time
intervals from roughly $2\,t_*$ before the peak to $2\,t_*$ after
the peak\footnote{ Explicitly, $t_i=t_0+[(4i-3)/n-2]\,t_*\,$,
$i=1,\ldots n\,$. This distribution is slightly asymmetric around the
peak.}. The lens positions are given by
\beq
l_i=\sqrt{p^2+[(t_i-t_0)/t_*]^2}
\label{eq:lenspos}
\eeq
for $i=1,\ldots n$. We compute the source star fluxes $F_*(l_i)$
directly from equation~(\ref{eq:main}) using one of the model
brightness profiles (not its approximation) and the
angle-integrated amplification from equation~(\ref{eq:aiampl}). We
integrate across the transit point divergence using
expression~(\ref{eq:divergence}) and, to match the accuracy of
this expansion, a local linear approximation of the brightness
profile. To obtain the simulated fluxes $F_i$ we add the blend
contribution and Gaussian noise $\eta_i$ with standard deviation
$\eta$ (as in the previous section),
\beq
F_i=[\,F_*(l_i)+\beta\,F_*(\infty)\,]\;[\,1+\eta_i\,] \quad .
\eeq
The errors are then given by $\sigma_i=\eta\,
[F_*(l_i)+\beta\,F_*(\infty)]$.

The studied parameter space is a Cartesian grid defined by all
combinations of values $\epsilon\,\in\{1,2,5,10\}$;
$p\,\in\{0,0.25,0.5,0.75\}$; $\beta\,\in\{0,0.25,0.5\}$;
$n_*\,\in\{10,20,40\}$; and $\eta\,\in\{0.005,0.01,0.02\}$. As shown
in Appendix A, higher values of $\epsilon$ provide
the same sensitivity to surface structure as $\epsilon=10$, even
though they provide higher overall amplification. For lenses with
Einstein radii lower than 1 the sensitivity drops, for even lower
values such events would be hard to detect at all, due to their
low amplification. Events with impact parameters $p\gtrsim1$
contain only a weak signature of the surface structure, so that
their inversion is useless under realistic noise conditions, as
shown in the previous section. The choice of blend parameters is
geared toward events with bright (e.g., red giant) source stars,
for which the fraction of blended light can be expected to be low.
The point numbers are representative and realistic. The $0.5\%$
noise level has been achieved in several microlensing events, the
two other values are more usual.

The model fitted to the light curves is based on
equation~(\ref{eq:decompflux}) with additional blended flux $F_B$,
\beq
F_M\,(l_i)=F_B+\alpha_1\,\mathbb{C}_{i1}(\,\epsilon\,,\,l_i\,)
+\alpha_2\,\mathbb{C}_{i2}(\,\epsilon\,,\,l_i\,)\quad ,
\label{eq:fitmodel}
\eeq
where the matrix elements
$\mathbb{C}_{ij}$ are given by equation~(\ref{eq:decmatrix}), and
the lens positions by equation~(\ref{eq:lenspos}). In this model
the asymptotic flux (flux in absence of the lens) from the lensed
star is $F_*(\infty)=\alpha_1\,C_{01} +\alpha_2\,C_{02}$, where
$C_{0j}=\lim_{l_i\to\infty}\mathbb{C}_{ij}$ for $j=1,2\,$. These
two values are $C_{01}\doteq 2.680\,,\,C_{02}\doteq 2.418$ for the
PCA basis and $C_{01}\doteq 3.142\,,\,C_{02}\doteq -1.047$ for the
LLD basis. The value $F_B/F_*(\infty)$ corresponds to the blend
parameter $\beta$ as defined above.

This seven-parameter model is linear in three parameters
($F_B,\,\alpha_1,\,\alpha_2$), and depends on the remaining four
($\epsilon,\,p,\,t_*,\,t_0$) through
$\mathbb{C}_{ij}[\,\epsilon\,,\,l_i\,(p,\,t_*,\,t_0)\,]$. In order to
speed up the subsequent computations, it is crucial to precalculate these
matrix elements for a sufficiently dense grid of Einstein radii $\epsilon$
and lens positions $l$ using the two basis functions ($f_{1,2}$ for PCA
and $g_{1,2}$ for LLD) and equations~(\ref{eq:aiampl}) and
(\ref{eq:divergence}), in a manner similar to the source light curve
computation described above. Values at intermediate points can be obtained
by interpolation.

Due to the form of the dependence on the nonlinear parameters the
full inversion problem cannot be solved  analytically. Instead we
find numerically the best-fit parameters by minimizing
\beq
\chi^2=\sum\limits_{i=1}\limits^{n}\sigma_i^{-2}\,
[\,F_i-F_M(l_i)\,]^2\quad .
\label{eq:fullchi}
\eeq
However, the numerical search can be limited to only four out of
the seven dimensions, due to the presence of three linear
parameters (as also noted by \citealt{rhie99} and
\citealt{albrow99b}). For any fixed combination of the four
nonlinear parameters, the combination of
($F_B,\,\alpha_1,\,\alpha_2$) minimizing the $\chi^2$ function can
be found analytically by solving a set of three linear equations.
These are the corresponding minimum conditions obtained by
differentiating equation~(\ref{eq:fullchi}). It is advantageous to
exclude non-physical solutions in advance by requiring for example
a non-negative blend, a non-negative brightness profile peaked at
the center (see Appendix B), and the total model asymptotic flux
being within $\pm\sigma_0$ of the baseline flux $F_0$ (both
constants estimated from the data). These five constraints
restrict the linear parameter space to a pentahedron. If the
computed minimum (for any fixed set of nonlinear parameters) lies outside
this region, the constrained minimum can be found on its surface.

For the minimum search in the remaining four-dimensional parameter
space ($\epsilon,\,p,\,t_*,\,t_0$) we used a Monte Carlo technique
based on gradual contraction of parameter intervals towards lower
values of $\chi^2$. This technique was found to be very robust in
finding the global minimum.

As the full inversion problem is nonlinear, we inverted 100 noisy
light curves for each point of the parameter grid to get a
realistic estimate of statistical errors. In order to find the
systematic error and bias of the method we tested the 32 model
brightness profiles listed in Appendix B. The distribution of the
values recovered from different noisy simulations was usually
found to be substantially non-Gaussian. Therefore we decided to
describe these distributions using the median instead of the mean,
and the narrowest interval containing $68.3\%$ of the recovered
values instead of the $\pm\sigma$ standard deviation interval,
which contains the same fraction of points in a Gaussian
distribution. \footnote{ These two descriptions converge in case
the underlying distribution is indeed Gaussian.}

With respect to the two photometry-defining parameters $\eta$ and
$n_*\,$, a good-quality fit not dominated by systematic errors is
indicated also by the linear scaling of the total error with light
curve noise and the inverse square root of the number of points
($\propto \eta\,n_*^{-1/2}$). We tested both these scalings; for
the $n_*$ scaling we used the results for $\eta=0.005$ in order to
suppress the effect of light curve noise.

\subsection{Recovered Lensing Parameters}

Before studying the recovered limb darkening, we check here the
accuracy of recovered lensing parameters: namely, Einstein
radius $\epsilon$, impact parameter $p$, blend parameter
$\beta=F_B/(\alpha_1\,C_{01}+\alpha_2\,C_{02}\,)$, source radius
crossing time $t_*$ and closest approach time $t_0$. For avoiding
confusion we add a subscript $`0'$ to denote the original values
used for the simulations, i.e., $\epsilon_0\,, p_0\,, \beta_0\,,
t_{*0}\,$ and $t_{00}$. For each point in the original parameter
grid, we statistically analyze the full set of $32\times100$
recovered parameter values.

We find that the best results for most parameters are obtained in
events with high Einstein radius, low impact parameter and low
blending. For example, the original parameter combination
$\{\epsilon_0=10; p_0=0; \beta_0=0; t_{*0}=1; t_{00}=0\}$ was
recovered using the PCA basis as
$\{\epsilon=10.11^{+0.12}_{-0.19}; p=0.016^{+0.011}_{-0.016};
\beta=(1.8^{+9.8}_{-1.8})\times 10^{-3};
t_{*}=0.9989^{+0.0073}_{-0.0068}; t_{0}=(-0.1^{+1.4}_{-1.6})\times
10^{-3}\}$ from simulated light curves with $n_*=40$ and
$\eta=0.01$. As an exception, the impact parameter is recovered
most accurately in events with higher impact parameters,
transiting closer to the limb of the source star. This is probably
due to the weak dependence of the microlensing amplification on
the brightness profile in the range $0.65\lesssim l\lesssim0.8$
(see Appendices A and B), which in turn implies a lower degeneracy
between the remaining fitted parameters (mainly $p$ and $t_*\,$).

The closest approach time $t_0$ is by far the best-recovered
parameter. The accuracy of the recovered Einstein radius and blend
parameter decreases slightly in events with $\epsilon_0\gtrsim5$.
In such events, better accuracy is obtained by including more
distant light curve points in the analysis, as discussed further
in \S 5.4. The total error of the PCA results is
$\propto\eta\,n_*^{-1/2}$ in events with $\epsilon_0\gtrsim1$. The
only found exceptions are the impact parameter error when $p_0=0$,
and the blend parameter error when $\beta_0=0$ (both are values at
the parameter space boundary).

The LLD results are in most cases qualitatively similar. In a few
cases the accuracy is slightly worse and the error scaling with
light curve noise and number of points does not follow
$\eta\,n_*^{-1/2}$ as accurately as in the PCA analysis. Clear
systematic differences can be seen in the recovery of the source
radius crossing time $t_*$. The most interesting difference is
found in the general scaling of the errorbars: the errors scale
roughly as $\eta^{0.7}$ and $n_*^{-0.3}$, quite consistently. A
closer study of the numerical results reveals that systematic
errors in the $t_*$ values recovered using LLD are large --
comparable to the statistical errors, hence the different
scalings. This means that while the LLD errors for $n_*=40$ and
$\eta=0.01$ are only slightly larger than the PCA errors,
decreasing the light curve noise or increasing the number of
points will only boost the relative advantage of PCA analysis over
LLD analysis. This problem is most probably caused by LLD
generally providing a poor fit to realistic brightness profiles,
as demonstrated in Appendix B.

\subsection{Recovered Limb Darkening}

In this subsection we first present the results obtained for three
sample input brightness profiles. We then proceed to estimate the
actual error of the reconstructed profile for a source with
unknown limb darkening by a combined analysis of the simulation
results for all 32 tested profiles.

The shape of the recovered profile is given by the PCA or LLD limb
darkening parameter $\kappa=\alpha_2\,/\alpha_1\,$. The accuracy
of recovering this parameter can be studied similarly as the
lensing parameters in the previous subsection. Of the 32 input
model brightness profiles (see Appendix B and
Table~\ref{tab:fits}), we present in Figure~\ref{fig:rld} the
results for the most centrally peaked profile (B3750g10), an
intermediate one (R3500g05), and the flattest profile (I4000g10).
We determine the median and the $68.3\%$ confidence interval from
the 100 noisy light curves computed for each input parameter
combination. In addition we plot the minimum and maximum obtained
values (none of the simulation results were discarded). The
results shown in the figure correspond to $n_*=40$ and
$\eta=0.01\,$.

As seen from the results for all profiles and both PCA and LLD
fits, limb darkening parameters are most accurately obtained for
events with low impact parameter $p_0$, although the dependence is
weak for $p_0\lesssim0.5$. The accuracy increases only very weakly
with higher Einstein radius $\epsilon_0$, and is fully independent
of blend parameter $\beta_0$. These results are encouraging --
either limb darkening parameter can be recovered even if the blend
or Einstein radius are poorly fitted.

In the case of profile B3750g10, the original LLD parameter value (1.0)
lies exactly at the parameter space boundary, as reflected also by the
recovered values. The plotted PCA errorbars for the ideal lensing
parameter combination ($\epsilon_0=10\,;\,p_0=0\,;\,\beta_0=0)$ are
consistent with best-fit PCA limb darkening parameters corresponding to
all the $B$ profiles in Table~\ref{tab:fits}. In the LLD case, the
errorbars are formally consistent only with the B3750g10 value, due to
their nearly zero width. However, this does not mean that the actual
profile is recovered better (see Table~\ref{tab:fits}).

The results for profile R3500g05 exhibit a nonlinear dependence of the LLD
errors on the light curve noise $\eta$, indicating non-negligible
systematic errors. The LLD medians are clearly correlated with the impact
parameter $p_0$. The PCA errorbars for the ideal lensing parameter
combination are consistent with best-fit PCA limb darkening parameters
corresponding to the R3500 and V4000 profiles in Table~\ref{tab:fits}. In
the LLD case, the errorbars are consistent with all profiles from R3500 to
B3500, including all $V$ profiles.

The recovered LLD medians for profile I4000g10 systematically
underestimate the original value. The PCA errorbars for the ideal
lensing parameter combination are consistent with best-fit PCA
limb darkening parameters corresponding only to the I4000 profiles
in Table~\ref{tab:fits}. The same result is obtained in the LLD
case. However, note here that no flatter profiles were used (see
Fig.~\ref{fig:profiles} and Table~\ref{tab:fits}).

From the general analysis we conclude that the PCA results have the
advantage of being uncorrelated with the lensing parameters and having no
apparent systematic problems. The LLD medians are correlated with $p_0$ and
sometimes with $\beta_0\,$; the results suffer from systematic errors due
to the inadequacy of the LLD approximation. Even in ideal cases, the LLD
results are less accurate than the PCA results.

It is worth noting that this test only told us how well a fitted
parameter of the actual brightness profile can be recovered from
the light curve. The conclusions confirm that the parameter of a
worse-fitting model will be recovered worse. The following part of
this subsection studies how well is the actual brightness profile
of the source recovered.

Using the results for all 32 profiles it is now possible to make an
estimate of the accuracy of the inverted brightness profile $\hat{B}(r)$,
given by equation~(\ref{eq:decomp}), for any given combination of input
parameters $(\epsilon_0\,,\,p_0\,,\,\beta_0\,,\,n_*\,,\,\eta)$ independent
of the actual source brightness profile. First we compute the relative
brightness profile deviation for each of the 32$\times$100 inverted light
curves,
\beq
\delta(r)=\;\frac{B(r)}{F_{*0}(\infty)}\;\frac{F_*(\infty)}{\hat{B}(r)}\,
-1\;=\;\frac{B(r)}{F_{*0}(\infty)}\;
\frac{\alpha_1\,C_{01}+\alpha_2\,C_{02}}{\alpha_1\,f_1(r)+\alpha_2\,f_2(r)}
\,-1\quad,
\label{eq:brdev}
\eeq
here $F_{*0}(\infty)$ is the asymptotic flux of the input model
profile $B(r)$ \footnote{The obtained profiles $\hat{B}(r)$ are normalized
to the total source flux in order to avoid uncertainties
introduced by any potential inaccuracy of the fitted blend.}. From
this full set of data we determine the total median $\delta_B(r)$,
which represents the bias of the inversion, and the $68.3\%$
confidence interval $(\delta_B-\delta_L\,,
\,\delta_B+\delta_U)\,$, where $\delta_L(r)+\delta_U(r)$ is the
total error.

To estimate the systematic error, we determine the median deviation for
each model profile separately, from its 100 noisy light curves. Analysis
of the distribution of these 32 profile-specific medians yields a
``median median" $\delta_{MM}(r)$, and the $68.3\%$ confidence interval
$(\delta_{MM}-\delta_{SL}\,,\,\delta_{MM}+\delta_{SU})\,$. The value
$\delta_{SL}(r)+\delta_{SU}(r)$ is the systematic error, a measure of
source-dependence of the inversion accuracy.

From the previous analysis we obtain a best estimate of the brightness
profile of the lensed star,
\beq
\mathcal{B}(r)=\left[1+\delta_B(r) \begin{array}{c} +\delta_U(r) \\
-\delta_L(r)\end{array}\,\right]\;\times\,\hat{B}(r) \quad .
\label{eq:brbest}
\eeq
If the accuracy of the recovered blend parameter is low, by normalizing
this result we still obtain the best estimate of the normalized profile of
the star.

A reliable inversion method requires not only a small total error
$\delta_L+\delta_U$, but also negligible bias $\delta_B/
(\delta_L+\delta_U)\rightarrow 0$, so that the inversion result is
closest to the best estimate. It is also desirable to have small
systematic errors $(\delta_{SL}+\delta_{SU})/
(\delta_L+\delta_U)\ll 1\,$, so that the inversion accuracy
doesn't depend substantially on the source profile, and the
inversion can yield better results with improved photometry. Such
a method would also have $\delta_{MM}\simeq\delta_B$.

Representative PCA and LLD results of the described analysis are
presented in Figure~\ref{fig:deve10} for $\epsilon_0=10$,
$\beta_0=0$, $n_*=40$ and $\eta=0.01$. The plotted curves include
the bias $\delta_B$, the total error as the 68.3\% confidence
interval $(\delta_B-\delta_L\,, \,\delta_B+\delta_U)\,$, the
``median median" $\delta_{MM}$ and the systematic error as the
68.3\% confidence interval
$(\delta_{MM}-\delta_{SL}\,,\,\delta_{MM}+\delta_{SU})\,$; all are
plotted as a function of the position on the stellar disk $r$. The
jagged shape of some of the curves is merely due to using only 32
model profiles in the analysis. All the desirable properties
mentioned in the previous paragraph can be achieved by PCA
inversion, but not by LLD inversion. The PCA results show a
negligible bias all the way to the limb of the source, the
coincidence of medians $\delta_{MM}\simeq\delta_B$, and systematic
errors much smaller than total errors -- with a single exception
in the case of the best-fit point at $r\simeq0.7$ with an error on
the order of 0.5\% \footnote{ This point occurs in the results
merely due to the fact that the intersection point of the PCA limb
darkening model (as discussed in Appendices B and A) lies in the
narrow intersection region of the used set of model profiles. This
is not the case for the LLD model.}. The quality of the inversion
improves weakly with higher Einstein radius $\epsilon_0$, and is
practically independent of the blend $\beta_0$ (only in poorly
fitted events the stability deteriorates very weakly with
increasing $\beta_0$). The errors scale linearly with light curve
noise $\eta$ and decrease as the inverse square root of the number
of points $n_*$ per source diameter crossing time. The best
inversion is achieved for low impact parameter events. However,
even at $p_0=0.75$ good accuracy can be achieved by obtaining a
higher number of good quality transit points. The results in
Figure~\ref{fig:deve10} demonstrate the best obtained inversion
(for $p_0=0$) has errors $\lesssim2\%$ from the disk center out to
$r\simeq0.9$. These errors can be further reduced by having better
than 1\% photometry or more than 40 light curve points per source
diameter crossing time.

While the LLD inversion results are not bad either, they do not achieve
the quality of the PCA results and do not have the desirable inversion
properties. The results are always biased, on average producing flatter
profiles. The bias can reach as much as 2\% in the region $r<0.9$ and
diverges further closer to the limb. The systematic errors are large and
often dominate the total error, particularly in the region $0.5\lesssim r
\lesssim 0.8\,$. The two medians $\delta_{MM}$ and $\delta_B$ differ by as
much as 1\%. The trends of the results with $\epsilon_0$, $p_0$ and
$\beta_0$ are the same as in the PCA case. The linear scaling of the
errors with $\eta$ and their inverse square root scaling with $n_*$ occurs
in fewer cases and to a lesser extent than in PCA inversion. The quality
of the achievable results is limited by the relatively large systematic
errors.

\subsection{Inversion of Observed Light Curves}

In the simulated inversions performed in the previous subsections
we only used points near the light curve peak, within $\pm 2$ {\em
source radius} crossing times from the closest approach. Light
curves of actual microlensing events usually provide data going
well beyond this region. To study the effect of more distant
points on the recovery of lensing parameters and the limb
darkening, we performed simulations including light curve points
within $\pm 2$~{\em Einstein radius} crossing times from the
closest approach.

The results for $\epsilon_0\gtrsim 5$ show a dramatic improvement
in the recovered Einstein radii and blend parameters, as mentioned
above. In such events impact parameters are recovered better as
well, with the exception of the already well recovered
$p_0=0.75\,$ which improves only marginally. The distributions of
the recovered source radius crossing times have more accurate
medians, but the total errorbars remain unchanged. On the other
hand, the accuracy of the recovered closest approach time and
(more notably) the limb darkening parameter is worse than while
fitting only the event peak. Consequently, the total error of the
recovered brightness profile at source center increases by 1--3\%.

These findings are a natural result of the light curve fitting procedure.
While fitting an extended light curve going far beyond the transit region,
the $\chi^2$ of the fit is altered more by a minute change in the global
parameters (e.g., $\epsilon$ and $\beta$) than by a more significant
change in the limb darkening parameter, which affects only points in the
transit region -- a small fraction of the total number of light curve
points.

High accuracy for the limb darkening of the source {\em as well
as} the lensing parameters can be achieved by a two-step
procedure. In the first step we fit the full light curve, varying
all parameters. In the second we fit only the transit region (not
more than $\pm 2$ source radius crossing times from the peak),
varying the limb darkening parameter and the closest approach
time, keeping the other parameters fixed at values obtained in the
previous step. Finally, we check the $\chi^2$ value of the overall
result.

\section{ANALYSIS OF MACHO ALERT 95-30 LIGHT CURVES}

In this section we analyze the observed light curves of the
microlensing source-transit event MACHO Alert 95-30
\citep{alc97d}. Altogether seven light curves were measured by the
MACHO and Global Microlensing Alert Network (GMAN;
\citealt{becker97}) collaborations. The individual light curves
were obtained by the MACHO team at Mount Stromlo Observatory (MSO;
MACHO B filter - 293 points used, MACHO R filter - 205 points),
and by the GMAN collaboration at Cerro Tololo Inter-American
Observatory (hereafter CTIO; R - 106 points), Mount John
University Observatory (MJUO; R - 41 points), University of
Toronto Southern Observatory (UTSO; R - 133 points, V - 127
points) and Wise Observatory (WISE; R - 17 points). \citet{alc97d}
presented the results of light curve fits using a uniform source
model as well as a linearly limb-darkened source with fixed limb
darkening parameters based on theoretical expectations. Here we
redo the general light curve fitting procedure\footnote{ The event
data were kindly provided by Andy Becker, who re-reduced the CTIO,
UTSO and WISE light curves recently.} to measure the limb
darkening profile of the source, as well as to obtain more
accurate lensing parameters.

The analysis can be performed in the manner used in the previous
section. The only difference is that this case involves a
simultaneous inversion of seven light curves. The four nonlinear
lensing parameters ($\epsilon,\,p,\,t_*,\,t_0$) are common to all
the curves, the linear parameters ($F_B,\,\alpha_1,\,\alpha_2$)
should be determined for each light curve separately. This is
necessary because of differences not only in the used filters, but
also in the optics, CCDs, point spread functions, data reduction
methods etc. Fitting the 25 parameters to the full set of data is
straightforward. As described in the previous section, for a given
combination of the four nonlinear parameters all the linear
parameters can be computed analytically by minimizing the total
$\chi^2$ function. Errors of the obtained best-fit parameters can
be obtained by Monte Carlo simulations. We generate 100 sets of
seven noisy light curves using the obtained best-fit parameters,
plus the timings and errorbars of the original data. We invert
these curves and from the distribution of the obtained values we
determine the (non-Gaussian) 68.3\% confidence intervals for each
parameter.

Estimation of the baseline parameters $F_0$ and $\sigma_0$ for the
two MACHO light curves revealed a systematic problem. As shown in
Figure~\ref{fig:machobase}, the baselines exhibit nearly periodic
correlated variations. There is an apparent period of $\sim45$
days, with an amplitude of $\sim3.5\%$ in MACHO B and $\sim2.9\%$
in MACHO R. Further inspection of the light curves suggests there
is additional irregular variability on both longer and shorter
timescales. Due to this complex variability, even within the span
of 10 days around the light curve peak the intrinsic flux from the
source may change by $\sim3\%$. As an immediate consequence, this
change will obscure the weak chromatic lensing effect close to the
source limb (see Appendix A). Moreover, the limb darkening
parameters (i.e., the source brightness profiles) can be expected
to vary, rendering any inversion questionable in principle.
Finally, the obtained lensing parameters will also be inaccurate
to a certain degree, due to the source variations in the light
curve peak region.

Proceeding formally with the fit, the problems become even more
apparent. The $\chi^2$ value for the fit to the original data is
substantially and systematically higher than the values obtained
for Monte Carlo simulated light curves with the best-fit
parameters. Even the residuals of the fit close to the peak show
systematic deviations. The relative residuals $F_i/F_M(l_i)-1$ are
plotted in Figure~\ref{fig:residpeak} for this part of the light
curve. As can be seen, the fairly large residuals in the different
light curves are clearly correlated (e.g., note the post-peak
changes).

The formal best fit parameters are: Einstein radius
$\;\epsilon=13.91_{-0.08}^{+0.13}\;$; impact parameter
$\;p=0.7253_{-0.0044}^{+0.0017}\;$; source radius crossing time
$\;t_*=2.5164_{-0.0028}^{+0.0060}\;d\;$ and the closest approach
time $\;t_0=1321.2359_{-0.0038}^{+0.0012}\;d\,$, measured from JD
2,448,623.50 \citep{alc97d}. Except for $t_0$, these values are
slightly different from those obtained by \citet{alc97d}. Partly
this is caused by the greater number of points near the light
curve peak in the current set of data (mainly in the UTSO light
curves), partly by the more general fitting procedure. The
greatest effect, however, is due to source variability. Neither of
the fitted models accounts for the effect of an intrinsically
irregularly variable source. All the lensing parameters obtained
here as well as in \citet{alc97d} should therefore be taken with a
grain of salt.

The values of the formal best-fit limb darkening parameters are
concentrated at the flat-profile boundary of the explored parameter
space. This can be seen also from the largely positive residuals
during the transit. Moreover, the best-fit limb darkening
parameters are mutually physically inconsistent (implying flatter
profiles in $V$ than in $R$ etc.). All these features confirm the
conclusion that limb darkening measurement from the MACHO Alert
95-30 event is not possible due to the variability of the lensed
star. It cannot be ruled out that some of the residuals close to
the peak may be due to the lens being binary
\citep{griesaf98,gausack00}, source being binary
\citep{griehu92,hangou97,hanjeo98} or even due to asymmetric
surface features on the source star \citep{hs00a,bh00}. However,
attempts to fit any of these models will also be largely hindered
by the variability of the source star.

It should come as no surprise that the lensed M4 giant source in MACHO
Alert 95-30 is variable. \citet{perpol98} state that ``virtually every M
giant is variable in brightness", with small-amplitude red variables being
two orders of magnitude more common than Miras. Variability larger than
1\% in the $V$ band sets in already in late K and early M type giants, and
increases further toward late M type giants \citep{joriss97,perpar98}. The
variability seems to be associated primarily with pulsations
\citep{perpar98,joriss97}, as suggested also by small-amplitude radial
velocity variations measured by \citet{cumm98}. The variations are not
strictly periodic, sometimes they are multi-modal or irregular
\citep{perpol98}. The periods can be on the order of tens of days down to
single days, as recently demonstrated from Hipparcos data by \citet{koel00}.

These results demonstrate the importance of checking the baseline
of source-transit event light curves to detect even
small-amplitude variability. Only events in which the change due
to intrinsic variability over the event timescale is sufficiently
smaller than the expected microlensing chromaticity (see Appendix
A) can be used for measuring the limb darkening of the source
star. As an exception, if the pulsations are well defined, regular
and stable, they can be accounted for a priori in the lensing
model, and the lensing effect can be fully separated from the
intrinsic variability. MACHO Alert 95-30 unfortunately does not
fall into this category of events.

\section{DISCUSSION}

The accuracy of measuring the limb darkening parameter from light
curves of eclipsing binaries was studied in detail by
\citet{popper84}. On the basis of light curve simulations Popper
concluded that achieving an absolute accuracy of $\pm0.1$ in the
linear limb darkening parameter requires on the order of 100 light
curve points within light curve minima obtained with 0.5\%
photometry. As shown in Figure~\ref{fig:rld}, analysis of
microlensing transit light curves frequently achieves better
accuracy even when using LLD inversion of light curves measured
with 1\% photometry. These results are surpassed in quality by PCA
inversion, which in fact directly demonstrates the inadequacy of
the linear limb darkening model. Going one step further, the
results in Figure~\ref{fig:deve10} show that PCA inversion can
retrieve realistic brightness profiles (instead of profile fits)
with an accuracy higher than 2\% from the center of the stellar
disk out to $r=0.9\,$. Inversion of microlensing light curves thus
gives the opportunity to proceed from measuring the limb darkening
parameter to measuring the actual limb darkening profile of the
source, thanks to the high achievable accuracy and low systematic
errors.

Eclipsing binaries have the advantage of having periodic light
curves, giving thus frequent opportunities to measure the eclipse
minima. Even though microlensing transit events are non-repeating,
giving only a single chance to obtain high quality photometry,
their analysis can place much better constraints on the limb
darkening as well as other properties of the source (HSL,
\citealt{hs00a}). Such events can also potentially provide results
for a broad range of stellar sources in the currently surveyed
Galactic bulge or Magellanic Cloud fields, as microlensing does
not depend on the stellar type of the source. The only biases are
introduced by the requirement of good photometry (giving an
advantage to brighter sources) and a sufficient number of transit
points (giving an advantage to longer transits, i.e., larger
sources).

The PCA method described in this work on the example of broadband
profiles of red giants can be simply extended to any other class
of sources, narrow-band filters etc. Using a set of brightness
profiles obtained from a corresponding class of atmosphere models
for the required wavelengths or passbands, an adequate PCA basis
can be constructed as described in \S 4.2. The first few terms
describing a sufficiently broad range of possible brightness
profiles can then be used for the light curve analysis.

Observed events with high quality photometry can also be used to
discriminate between classes of stellar atmosphere models with
different PCA bases. The more suitable class can be distinguished
mainly by the pattern of residuals of the light curve fit. The
best-fit model should then be checked whether it is in agreement
with measured properties of the unamplified source -- such as
colors, line strengths etc. Microlensing thus provides unique new
tests that will contribute to our understanding of stellar
atmospheres.

Microlensing transit events are much more frequent for caustics of
binary lenses than for single point lenses studied in this work.
The reason is that the probability of a caustic transit of the
source is usually given by the angular size of the caustic rather
than by the angular size of the source. While the source size
still determines the duration of the transit, detection of the
source entering a caustic gives an opportunity to predict and
observe in detail at least the exit from the caustic. Limb
darkening parameters have already been measured for four
caustic-crossing events in this way, three of them by The Probing
Lensing Anomalies NETwork (PLANET) collaboration and one by a
joint effort of five microlensing collaborations \citep{afonso00}.
The most recent one at the time of writing was described by
\citet{albrow01}. The sensitivity of a linear caustic-crossing
event to the brightness profile of the source was recently studied
by \citet{gaugou99} and \citet{rhieben99}. As shown by
\citet{rhieben99}, the chromaticity is generally lower than for a
single lens. On the other hand, a linear caustic always transits
the source completely (with ``zero impact parameter"). While
single lens transits have the potential to deliver even better
results, binary caustic-crossing events are already delivering
good results. The approach demonstrated here for the single lens
case can be readily applied to the binary case. The linearity of
the PCA approach will be particularly beneficial, as the fitting
of binary microlensing light curves is computationally heavily
burdened by proliferating nonlinear parameters
\citep{albrow99b,albrow00}. This method will substantially improve
the analysis of all previously reported as well as upcoming
caustic-crossing microlensing events.

\section{SUMMARY}

In this paper we show that limb darkening profiles of source stars
in microlensing transit events can be measured by light curve
inversion using brightness profile decomposition. Other inversion
techniques at present fail to reach the accuracy achieved by this
method.

Initial tests in \S 4.1 show that under the usual photometric
accuracy only two terms of the decomposition can be reasonably
recovered. Using linear limb darkening, which suffers from
systematic errors, is not ideal. It is necessary to find a limb
darkening basis providing in two terms a better description of
realistic brightness profiles than does linear limb darkening
(LLD). Such a basis can be constructed by principal component
analysis of a set of realistic model brightness profiles, as
demonstrated in \S4.2 . The obtained PCA basis provides the best
possible linear description of the used set of profiles, as
demonstrated in Appendix B. This method of describing limb
darkening should in fact replace the ill-suited and obsolete
standard power-law limb darkening model in other practical
applications as well.

We demonstrate the described inversion method in \S 5 by fully
analyzing simulated light curves. In \S 5.2 we check the accuracy
of recovering the basic lensing parameters. We show that best
results are obtained for events with Einstein radii
$\epsilon\gtrsim5$, low impact parameters $p<0.5$ and negligible
blend parameters $\beta\ll1$. As demonstrated in \S 5.3, the limb
darkening parameter is recovered more reliably using the PCA basis
than using the LLD basis, achieving also better brightness profile
resolution. The quality of these results is further demonstrated
by directly measuring the inversion accuracy, the difference
between the recovered and the original brightness profile. The
best results are obtained for small impact parameters $p<0.5$.
They are only very weakly dependent on the Einstein radius, and
independent of the blend. For events with small impact parameters
even light curves with 1\% photometry can yield an inversion
accuracy of $\pm2\%$ from the stellar disk center to $r=0.9$ near
the limb using the PCA basis. The accuracy can be further improved
by better photometry as well as by increasing the number of
measured transit light curve points.

The measured light curves of the microlensing transit event MACHO
Alert 95-30 \citep{alc97d} reveal the intrinsic variability of the
lensed red giant, as demonstrated in \S 6. The time scale,
amplitude and irregularity of the variability indicate that the
source changes significantly at the light curve peak, and thus its
limb darkening cannot be well constrained. This example
underscores the need to check the baseline of microlensing transit
events for signs of variability as low as 1\% on the event
timescale. Inversion results cannot be considered reliable without
this safety test.

Binary microlensing events, in which a caustic transits the source
star, are much more frequent than the single lens events studied
here. Though their sensitivity to the brightness profile of the
source tends to be lower, such events have already been used to
measure the limb darkening parameters of several stars. The
analysis of these caustic-crossing events will be significantly
improved by using the PCA approach to limb darkening as described
in this work.

\acknowledgments

It is a pleasure to thank Dimitar Sasselov for his enthusiastic
support, many helpful discussions and comments on the manuscript,
as well as for providing the model atmosphere data. The MACHO
Alert 95-30 event analysis was made possible by Andy Becker, who
kindly supplied the newly reduced light curves. I would like to thank
Ramesh Narayan for discussions about inversion techniques,
particularly for suggesting to use principal component analysis.
Finally I thank the referee, Andy Gould, whose comments and
suggestions helped to improve the manuscript.

\appendix

\section{BROADBAND LIGHT CURVE CHROMATICITY}

Following HSL, Figure~\ref{fig:lccomp} demonstrates the dependence
of the microlensing amplification on the surface brightness
profile of the source. In this example a red giant source
(effective temperature T=3750 K, surface gravity $\log{g}=0.5$) is
transited by an $\epsilon=10$ lens at zero impact parameter $p$.
$BV\!RI$ light curves of this event in the lower panel are to be
compared with the corresponding normalized brightness profiles in
the upper panel (see Appendix B for details). The microlensing
amplification clearly traces the shape of the brightness profile.
More centrally peaked profiles have more centrally peaked
amplification curves (light curves for $p=0$). Profiles which have
a relatively higher amplification at the source center mostly have
a relatively lower amplification close to the source limb, and
vice versa.

The sensitivity of microlensing to the brightness profile of the
source can be measured by the relative variation of amplification
with wavelength, due to the wavelength dependence of the profiles.
Following HSL, we define the microlensing chromaticity as the
ratio of the total amplification range to the average
amplification $[A_{max}-A_{min}]/ A_{aver}$ at a given lens
position. The maximum, minimum and average amplifications are
evaluated for a particular source over a given spectral range at a
given spectral resolution. HSL presented the chromaticity of low
and high resolution spectra, here we demonstrate the broadband
chromaticity using merely the four profiles from
Figure~\ref{fig:lccomp}. Plotted as a function of lens position in
Figure~\ref{fig:lcchrom}, the chromaticity curve has the same
character as those presented in HSL with a slightly lower
amplitude due to the lower resolution. The individual curves
correspond to lenses with different Einstein radii. Note that the
chromaticity becomes independent of $\epsilon$ for
$\epsilon\gtrsim5\,$. The highest sensitivity to limb darkening
occurs when the lens is close to the source center, another
somewhat less sensitive region is close to the source limb. The
sensitivity drops rapidly for non-transit lens positions. These
overall features were first reported in the works of
\citet{witt95,snw95,v-g95,bogcher95} and \citet{gouwel96}.

In the vicinity of $l\simeq0.73$ the amplification is largely
achromatic, i.e., independent of the underlying limb darkening
profile. The occurrence of this achromatic point is a direct
consequence of the narrowly confined region around $r\simeq0.69$
where all the normalized brightness profiles in the upper panel of
Figure~\ref{fig:lccomp} intersect. As Figure~\ref{fig:profiles}
and even the higher resolution results in HSL demonstrate, this
narrow intersection region is common to all the computed profiles
(see also Appendix B). It can be shown that this property is an
indication that the full range of profiles is well modelled by a
linear combination of two radial functions. Any such a suitable
model has an exact intersection point common to all the profiles.
The position of this point, however, depends on the particular
model. As a consequence, using a limb darkening model with a wrong
intersection point in the analysis of microlensing light curves
will additionally result in inaccurate values of the inferred
microlensing parameters. However, as mentioned in \S 7 such a
wrong model could be told by the pattern of residuals of the
light-curve fit. See Appendix B for more details on limb darkening
profiles.

\section{LIMB DARKENING PROFILES AND THEIR FITS}

In this work we used limb darkening profiles obtained from eight
model atmospheres of cool red giants computed by Dimitar Sasselov
(see HSL for details). The effective temperatures and surface
gravities of these models are T$\in\{$3500 K, 3750 K, 4000 K$\}$
and $\log{g}\in\{$0, 0.5, 1.0$\}$. Specific intensity was computed
for 17 rays passing through the atmosphere at different positions
on the projected disk. These positions are marked in
Figure~\ref{fig:profiles}. The $BV\!RI$ profiles were obtained
from these results using the filter response functions provided by
\citet{bessell90}. To obtain a full radial profile, these 17
points were interpolated using a cubic spline with boundary
conditions $B'(0)=0$ and $B''(1)=0$ \footnote{ The latter
condition is fairly arbitrary, but has no global effect on the
quality of the interpolation (see Fig.~\ref{fig:profiles}) due to
the high number of computed points constraining the profile close
to the limb.}. The variation of the profiles {\em in the studied
physical range} is strongest with wavelength (filter), weaker with
temperature and weakest with surface gravity. All profiles with
$\log{g}$=0.5 are plotted in Figure~\ref{fig:profiles}, normalized
to unit total flux. Two trends are obvious: the profiles at longer
wavelengths are flatter; so are mostly the profiles of higher
temperature models in the studied range.

The two limb darkening models used in this paper (PCA, LLD) are
demonstrated in Figure~\ref{fig:pcalldprof}. The PCA basis was
obtained directly by principal component analysis of all 32
computed model profiles, as described in \S 4.2. The standard
linear limb darkening (LLD) basis is defined by
equation~(\ref{eq:lld}). Figure~\ref{fig:pcalldprof} compares the
full ranges of profiles covered by the two models, normalized to
unit flux and plotted on the same scale as
Figure~\ref{fig:profiles}. The ranges are limited physically by
requiring positiveness and a central peak for the profile. Notice
that the LLD model is particularly ill-suited for describing the
peaked $B$ profiles in Figure~\ref{fig:profiles}, as well as the
behavior close to the limb in general. Also note that LLD profiles
with low values of the parameter $\kappa=\alpha_2\,/\alpha_1$ are
unrealistically flat. Finally, there is a difference in the
position of the intersection point as discussed in Appendix A. The
LLD point is $\gtrsim0.05$ closer to the limb than the PCA point,
which lies directly in the narrow intersection region of the
profiles in Figure~\ref{fig:profiles}. As a result, simulated
microlensing light curves generated using LLD will intersect at
$l\simeq0.77$, those generated using PCA at $l\simeq0.73\,$. This
is the probable cause for the large systematic error in the source
radius crossing time $t_*$ recovered from microlensing light
curves using LLD in \S 5.2.

The difference in the quality of least-squares fits\footnote{
Computed as unweighted fits to the spline-interpolated profiles
(see Fig.~\ref{fig:profiles}).} to all the 32 computed profiles
using the two limb darkening models is demonstrated by the results
in Table~\ref{tab:fits}. The columns of the table contain for both
limb darkening models the best-fit limb darkening parameter
$\kappa=\alpha_2\,/\alpha_1\,$, the normalized root-mean-square
deviation
of the fit,\\
\centerline{$\delta_{fit}=\{\int[B(r)-\alpha_1f_1(r)-\alpha_2f_2(r)]^2dr/
\int B^2(r)dr\}^{1/2}\;$,}
and the relative flux excess $\Delta F/F$ of the best-fit solution
over the original profile. The PCA fits always leave an r.m.s.
deviation $\delta_{fit}\lesssim0.7\%$, while the LLD fits can
leave a deviation of as much as 3\% (in the case of the B
profiles). The LLD fits are marginally better than the PCA fits
only for some of the R3750 and R4000 profiles. The more striking
difference is in the excess flux. While the fluxes of the PCA fits
differ from the actual source fluxes by less than
$2\times10^{-4}$, the LLD fits overestimate the source flux for
all B, V, R3500 and R3750 profiles, by as much as $0.75\%\,$. This
renders the LLD model also less useful for fitting microlensing
light curves, where such a difference is not negligible. In the
remaining profiles the flux is underestimated; a flux accuracy
comparable to the PCA fits is never achieved.

Although our PCA basis was constructed from limb darkening
profiles of cool giant models, Figure~\ref{fig:solarplot}
demonstrates this PCA model can be better suited than the LLD
model even for fitting measured data of narrow-band solar limb
darkening. For this figure we used data published by
\citet{mitchell59} spanning the $BV\!RI$ spectral range, namely
his measurements at 450, 545, 660 and 848~nm. The radial
coordinates of the data points are marked in the figure by
crosses. In order to give even weight to all radii\footnote{ Note
that the narrow region close to the limb has the highest density
of data points.}, we interpolated the observed data by a spline in
the manner described at the beginning of this appendix to obtain
$B(r)$. We performed least-squares fits to these interpolations
using both limb darkening models and plotted the deviation curves
$B(r)/B_{fit}(r)-1$. The small-amplitude bumps in the curves are
due to the interpolation. As seen from the figure, while for
450~nm and 848~nm both models have a comparable maximum deviation
(within the range $r\lesssim0.9$), for 545~nm and 660~nm the PCA
model fits the data better. It is interesting to note that here
the LLD model systematically provides fits more peaked than the
original profiles.

A more rigorous approach for studying different stellar types would be to
construct PCA bases from model atmospheres of such stars. The reason for
the cool giant PCA basis fitting the solar data is primarily physical. For
both stellar types as well as for a wide range of other cool stars, the
dominant source of opacity determining the limb darkening in the optical
range is the same --- namely, the H$^-$ ion (e.g., \citealt{gray92}).

\clearpage

\bef[t]
\epsscale{0.6}
\plotone{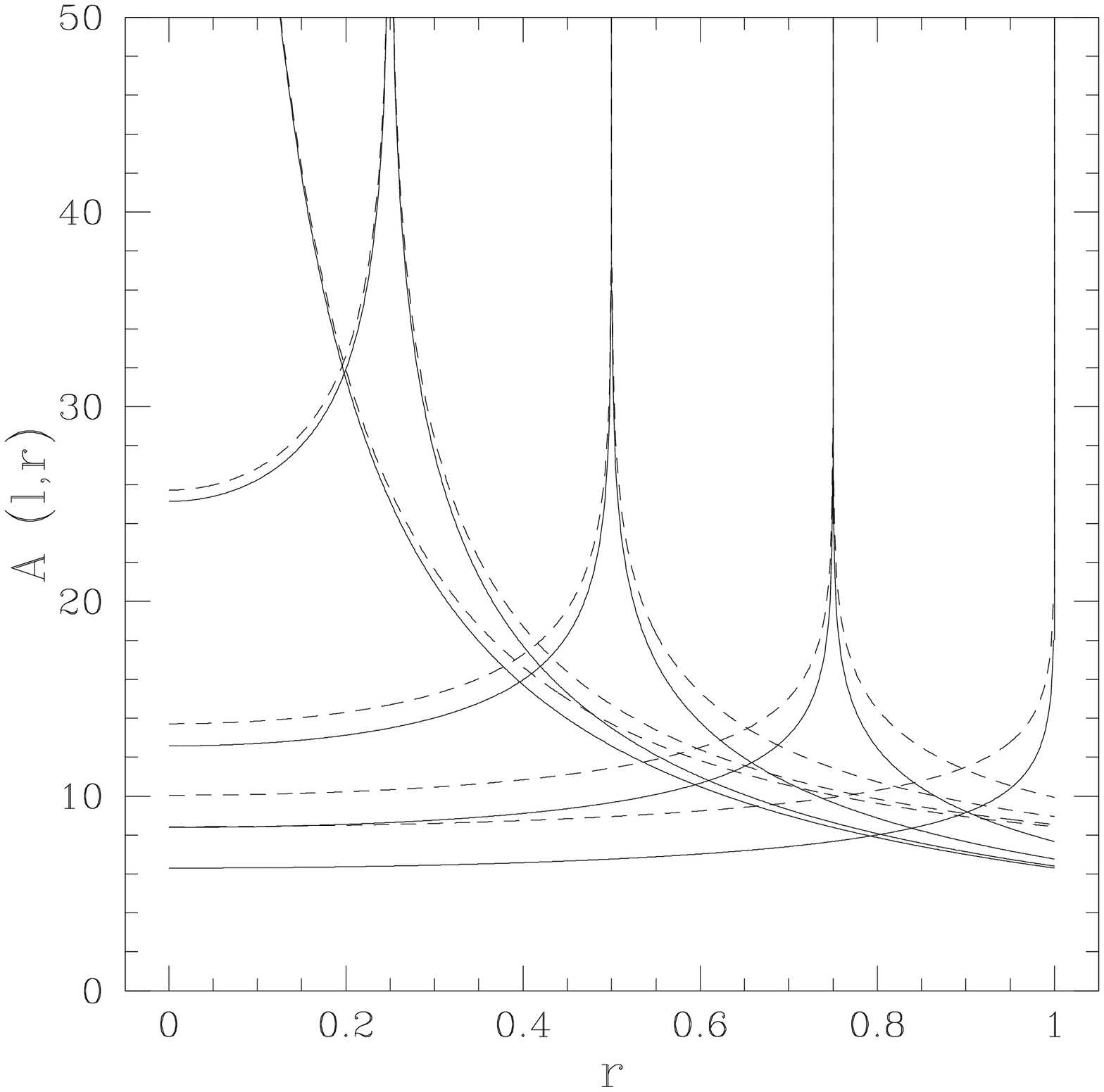}
\caption{Dashed line: Angle-integrated amplification $\mathcal{A}(l,r)$
for a lens with Einstein radius $\epsilon$=1 as a function of radial
distance $r$ for lens positions $l$=0, 0.25, 0.5, 0.75 and 1 (curves with
peaks from left to right). Solid line: $\mathcal{A}(l,r)\,/\,\epsilon\,$
for a lens with $\epsilon$=10, same lens positions.}
\epsscale{1}
\label{fig:aiampl}
\enf

\bef[t]
\epsscale{0.8}
\plotone{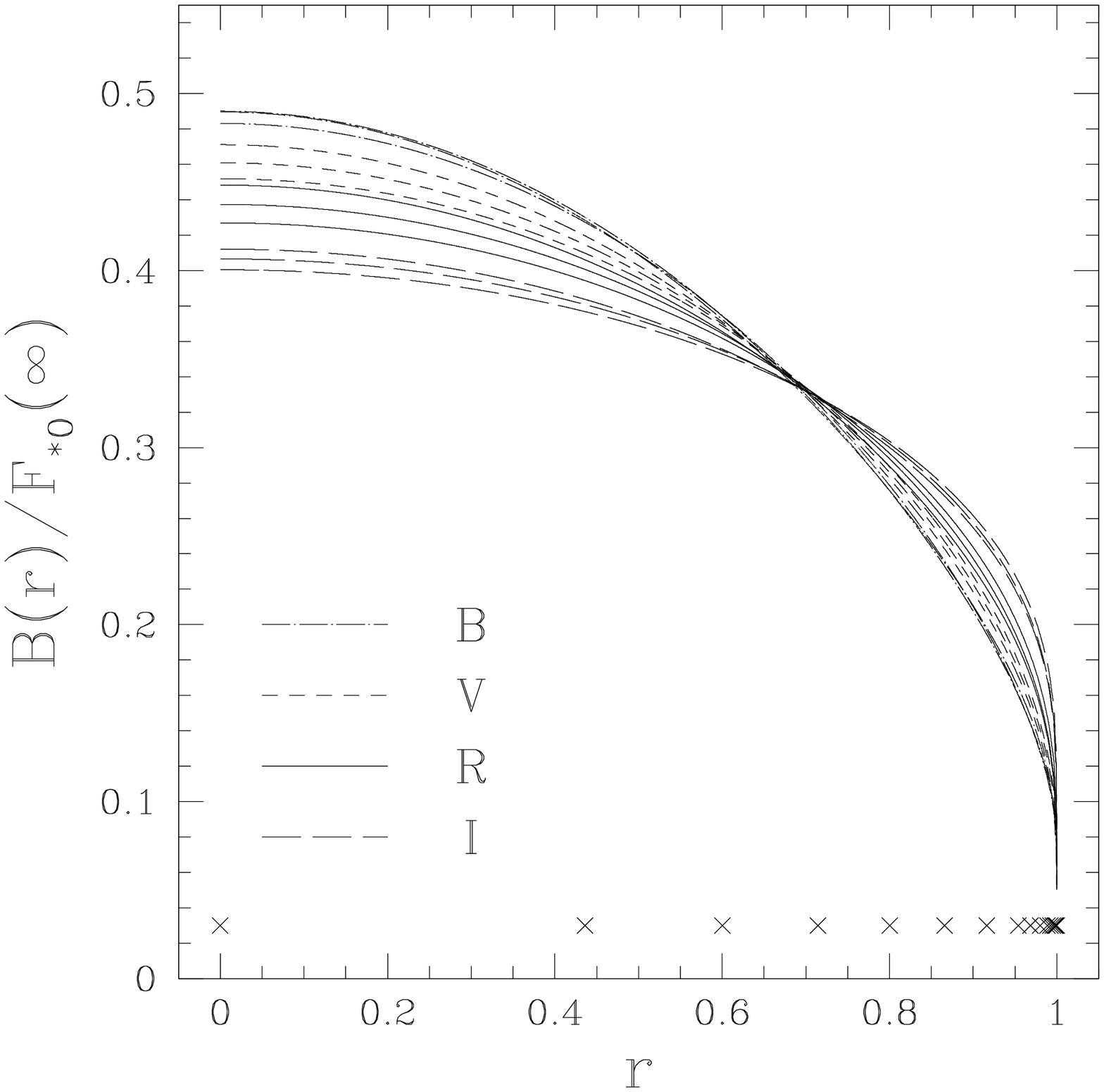}
\caption{Normalized $BV\!RI$ brightness profiles of three $\log{g}=0.5$
model atmospheres (from upper to lower at $r=0$ in each filter): T=3500,
3750, 4000 K. Radial positions of computed points are marked at
bottom, plotted curves are spline interpolations to these points. }
\label{fig:profiles}
\epsscale{1}
\enf

\bef[t]
\epsscale{0.5}
\plotone{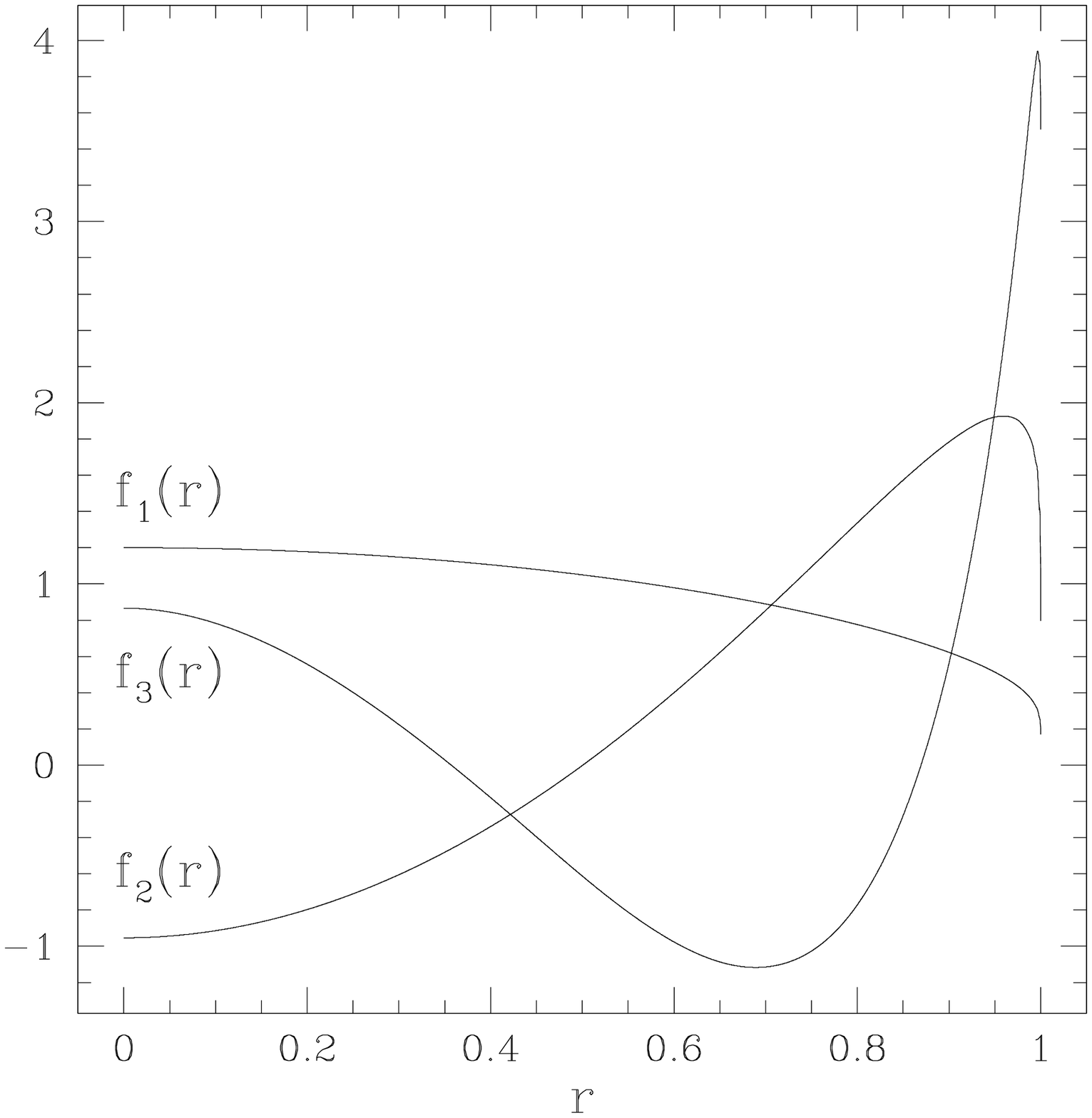}
\caption{First three orthonormal functions of the PCA basis for brightness
profile decomposition, obtained by principal component analysis of 32
brightness profiles of red giant model atmospheres (see
Table~\ref{tab:fits}).}
\epsscale{1}
\label{fig:basis}
\enf

\bef[t]
\centerline{~~~B3750g10~~~~~~~~~~~~~\hfil R3500g05
\hfil~~~~~~~~~~~~~I4000g10}
\plotthree{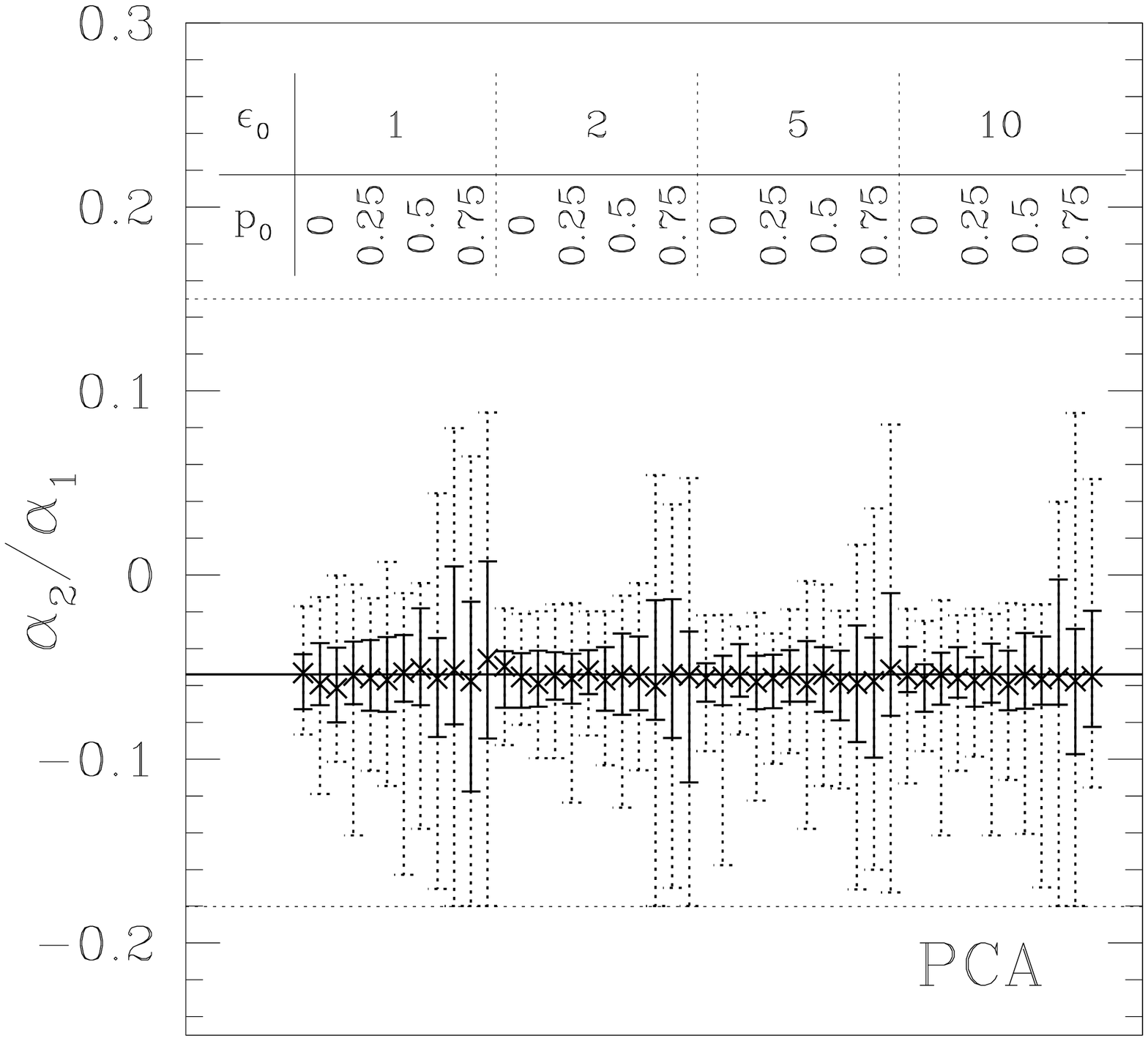}{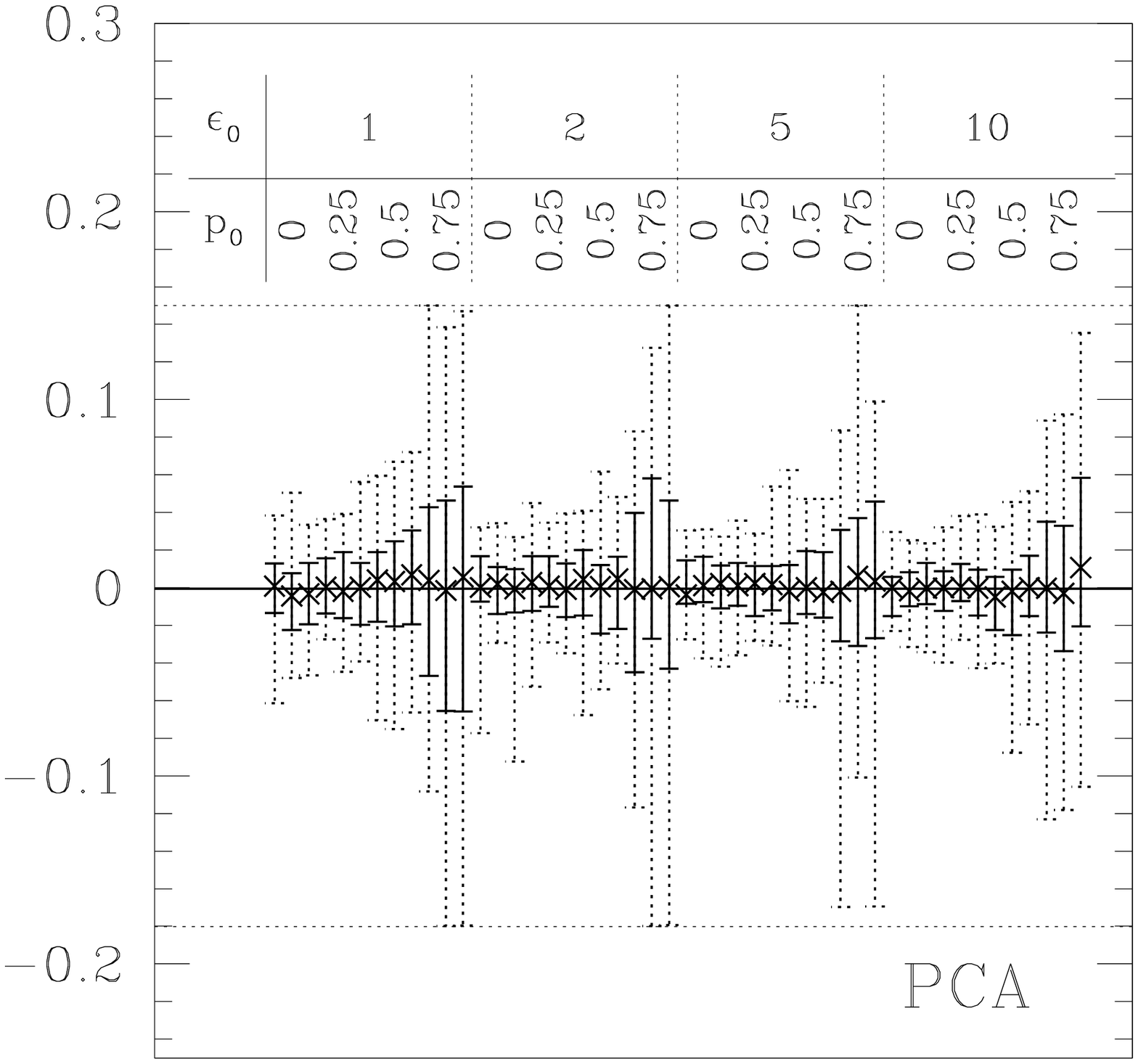}{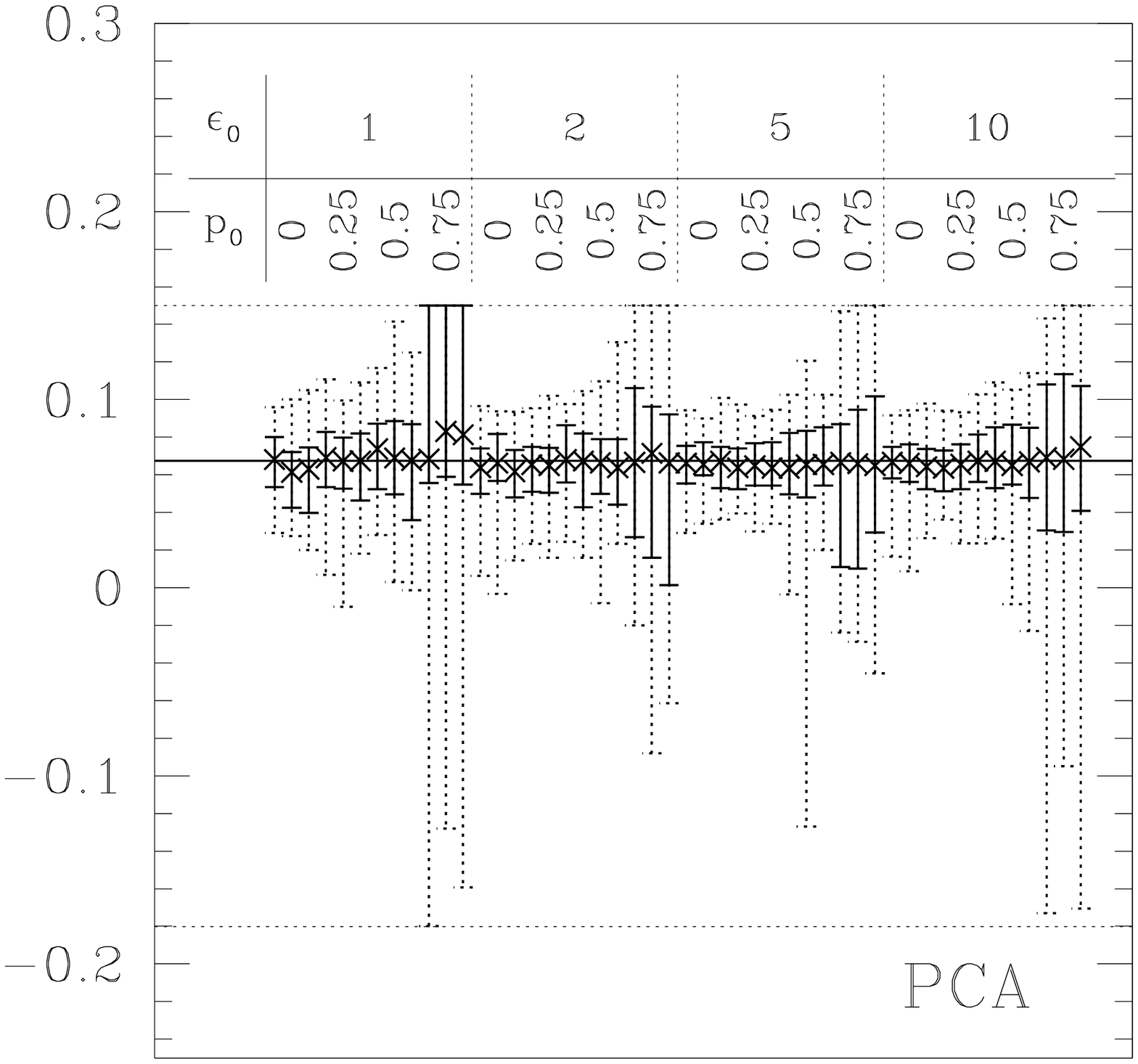}
\vspace{-0.2in}
\plotthree{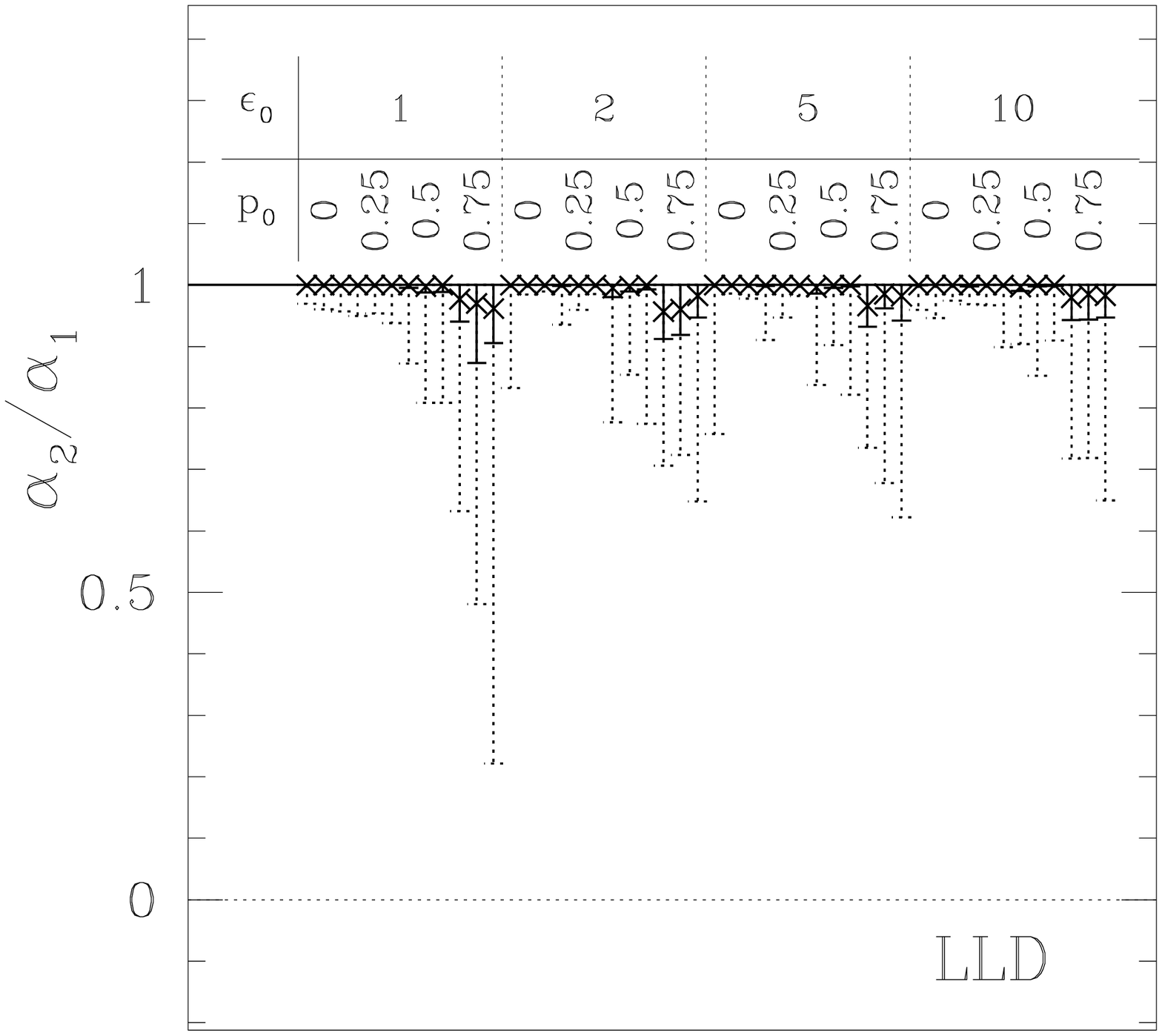}{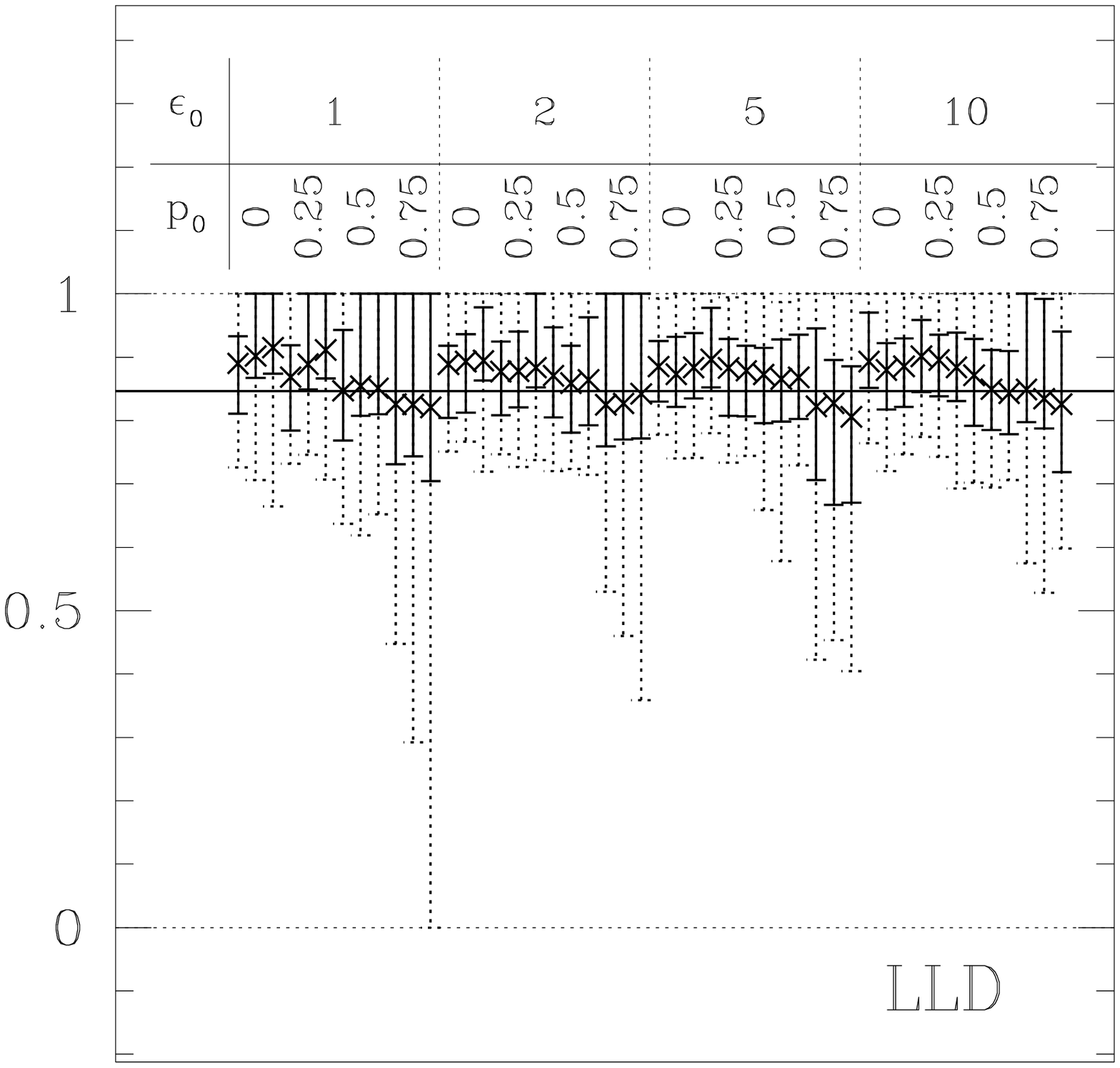}{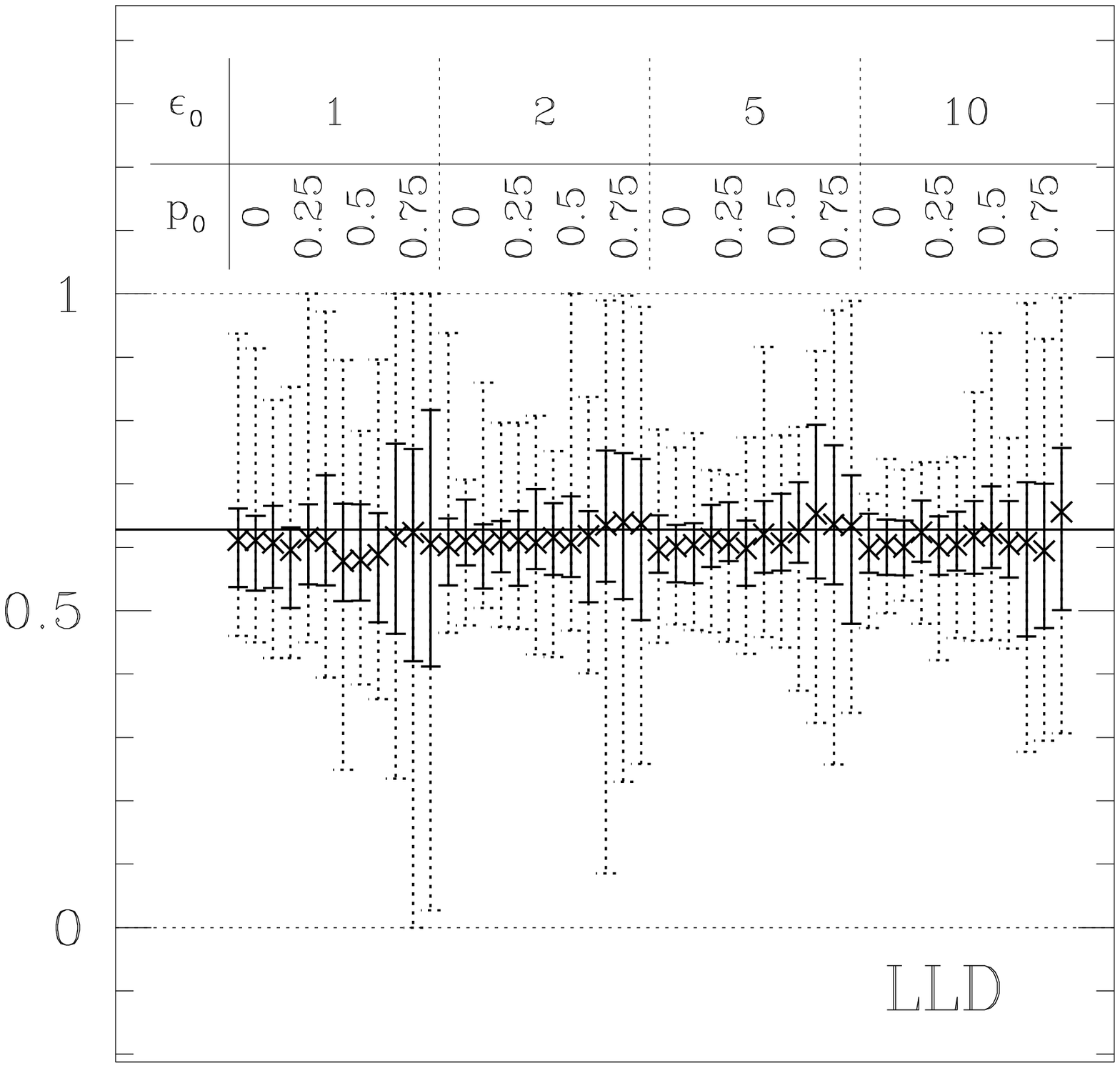}
\caption{Recovered PCA (upper panels) and LLD (lower panels) limb
darkening parameters for model profiles: B3750g10 (left column), R3500g05
(central column) and I4000g10 (right column). Individual points correspond
to values of the original Einstein radius $\epsilon_0$ and impact
parameter $p_0$ directly above them in the table, triplets of neighboring
points correspond to blend parameters $\beta_0=0,0.25,0.5\,$. Dotted
vertical bars: range of values from all simulations. Solid horizontal
lines: original best-fit values. Dotted horizontal lines: full parameter
range.}
\label{fig:rld}
\enf

\bef[t]
\ploteight{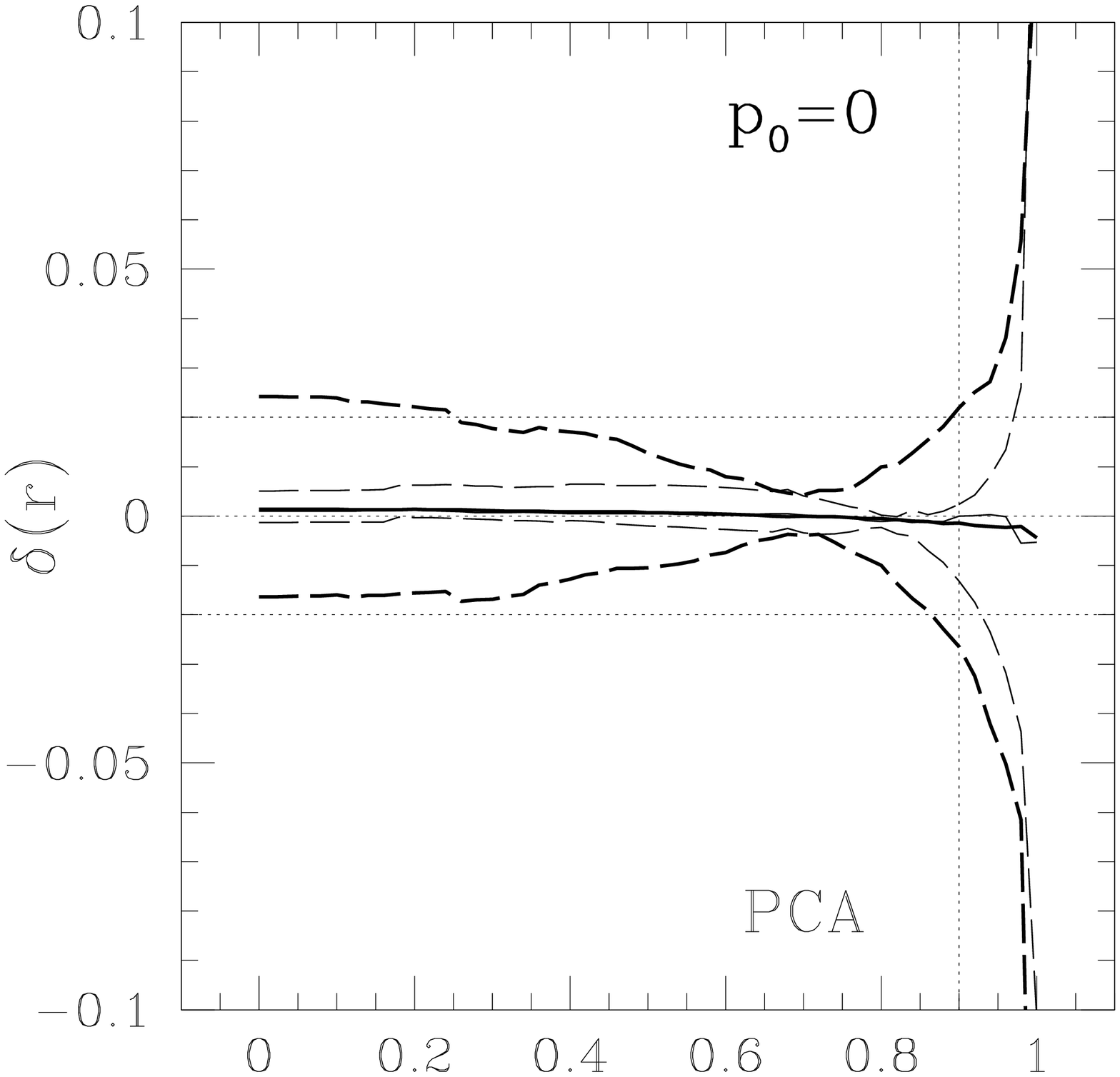}{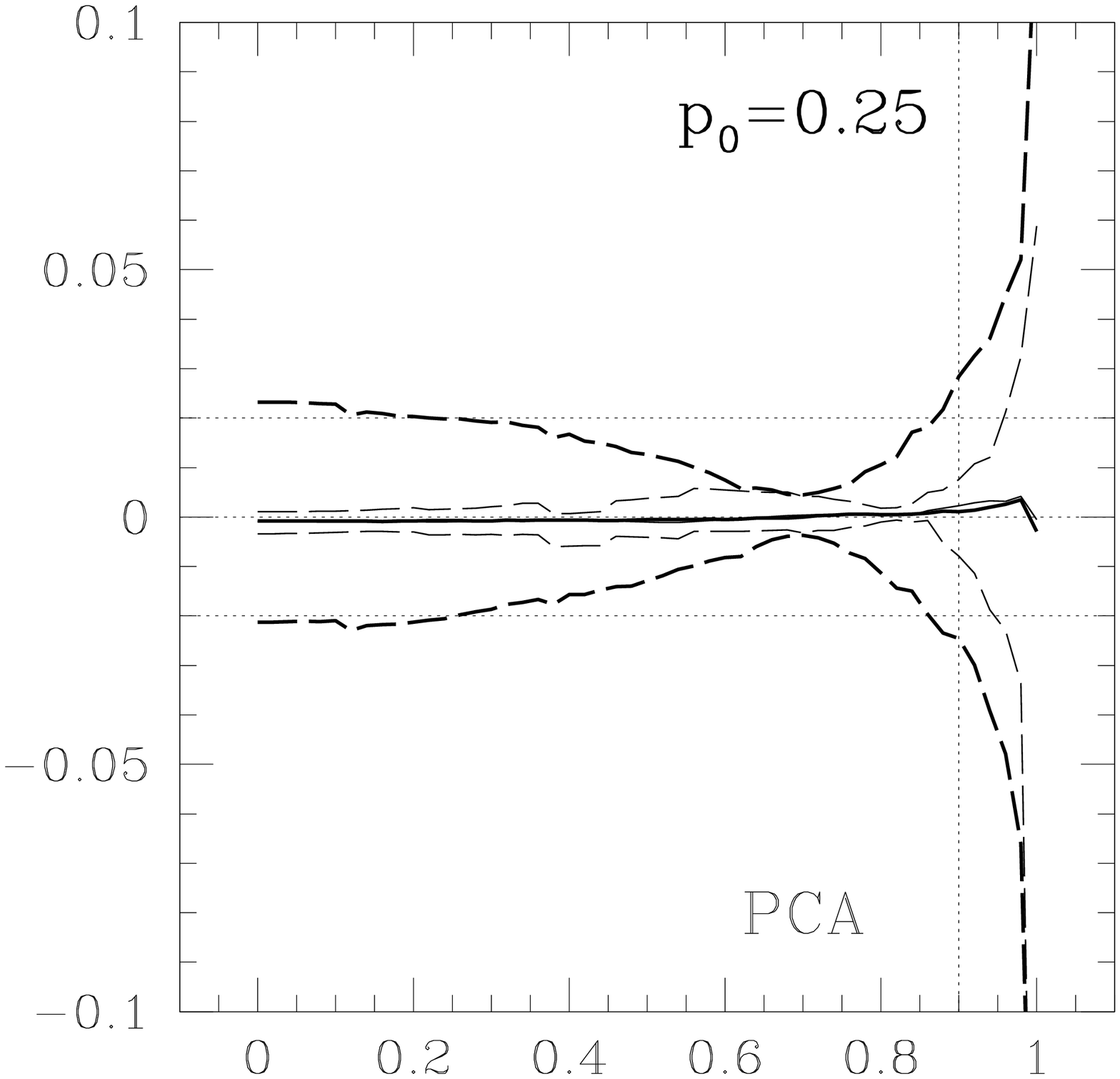}{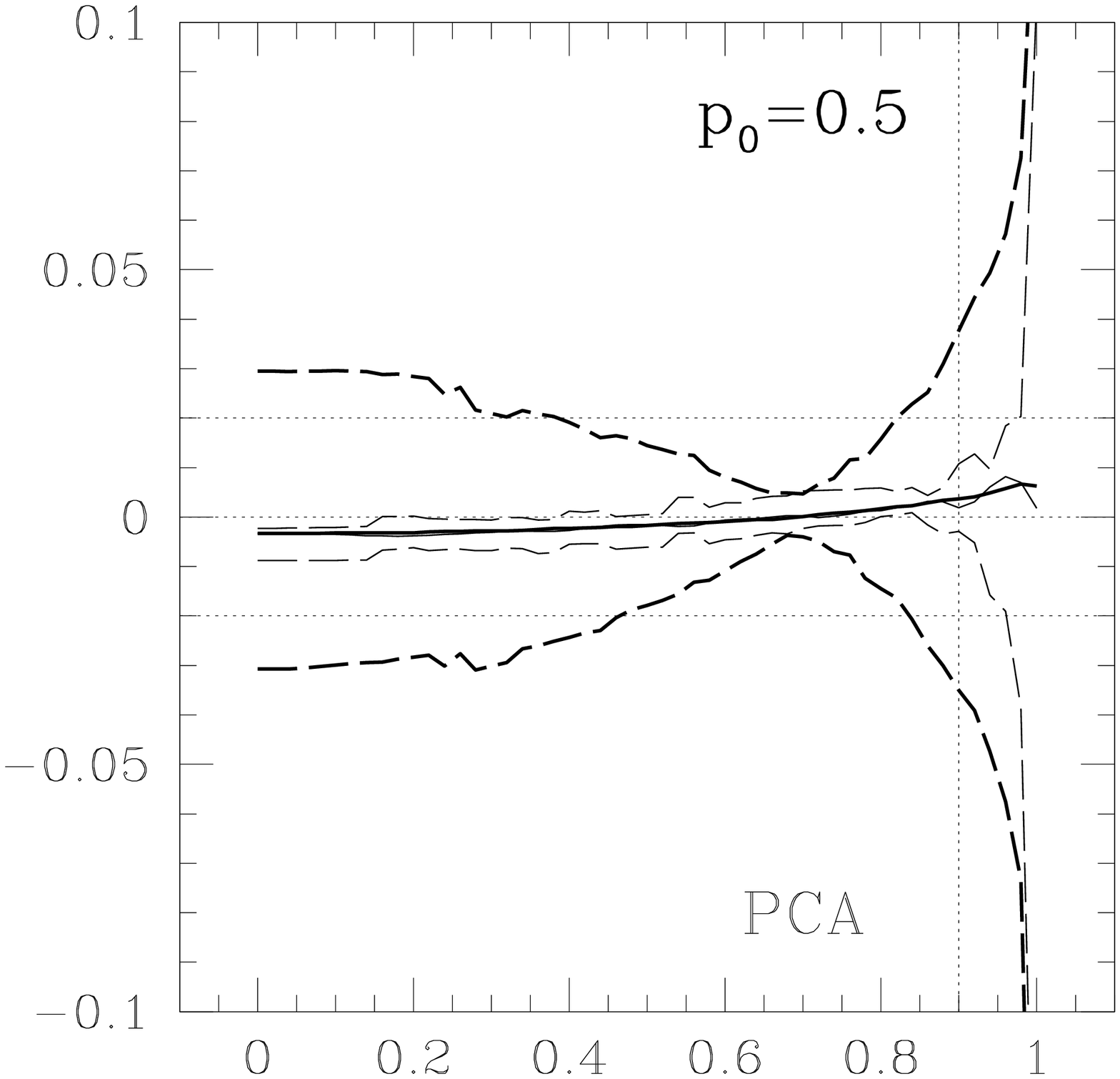}{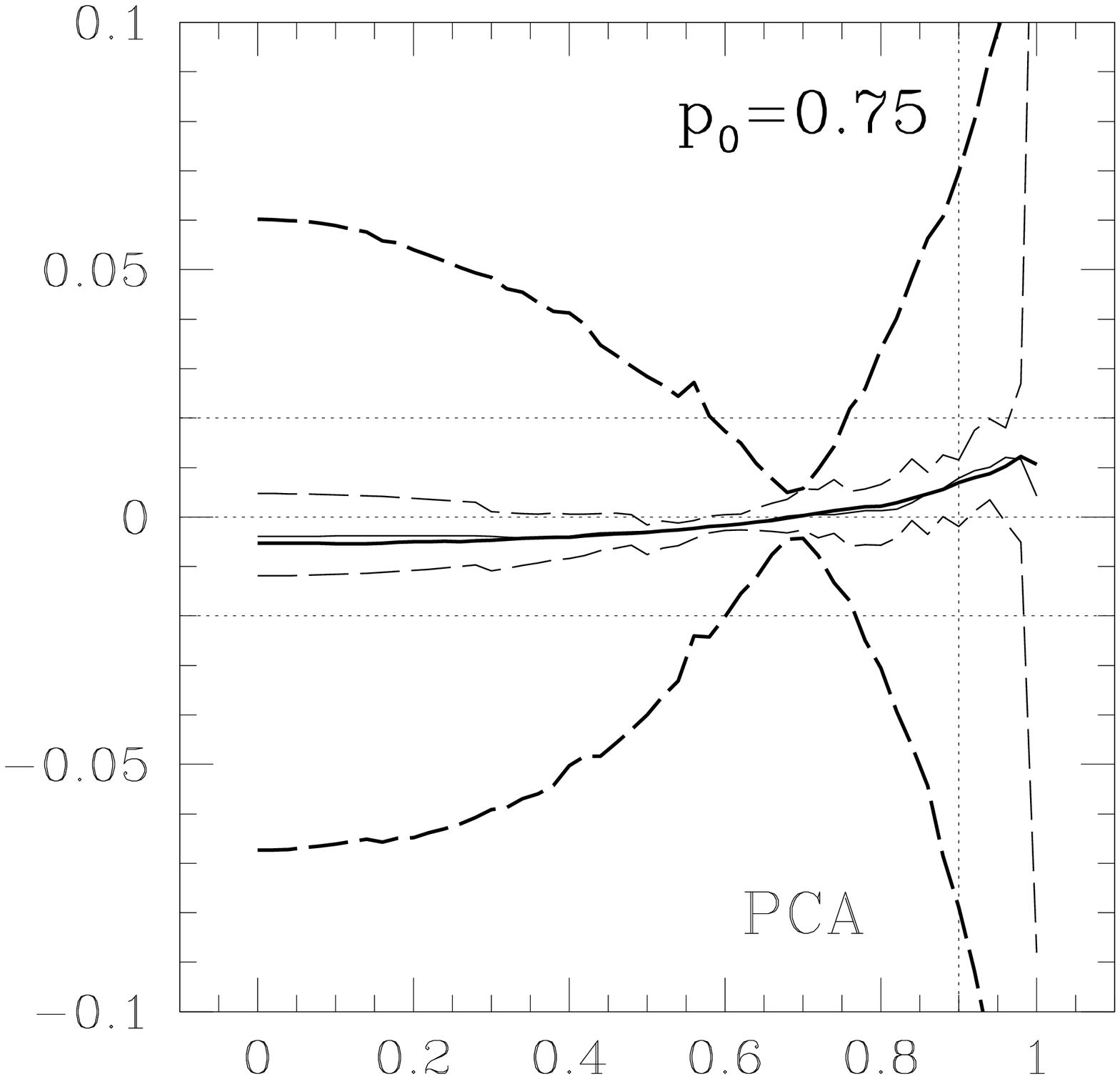}
{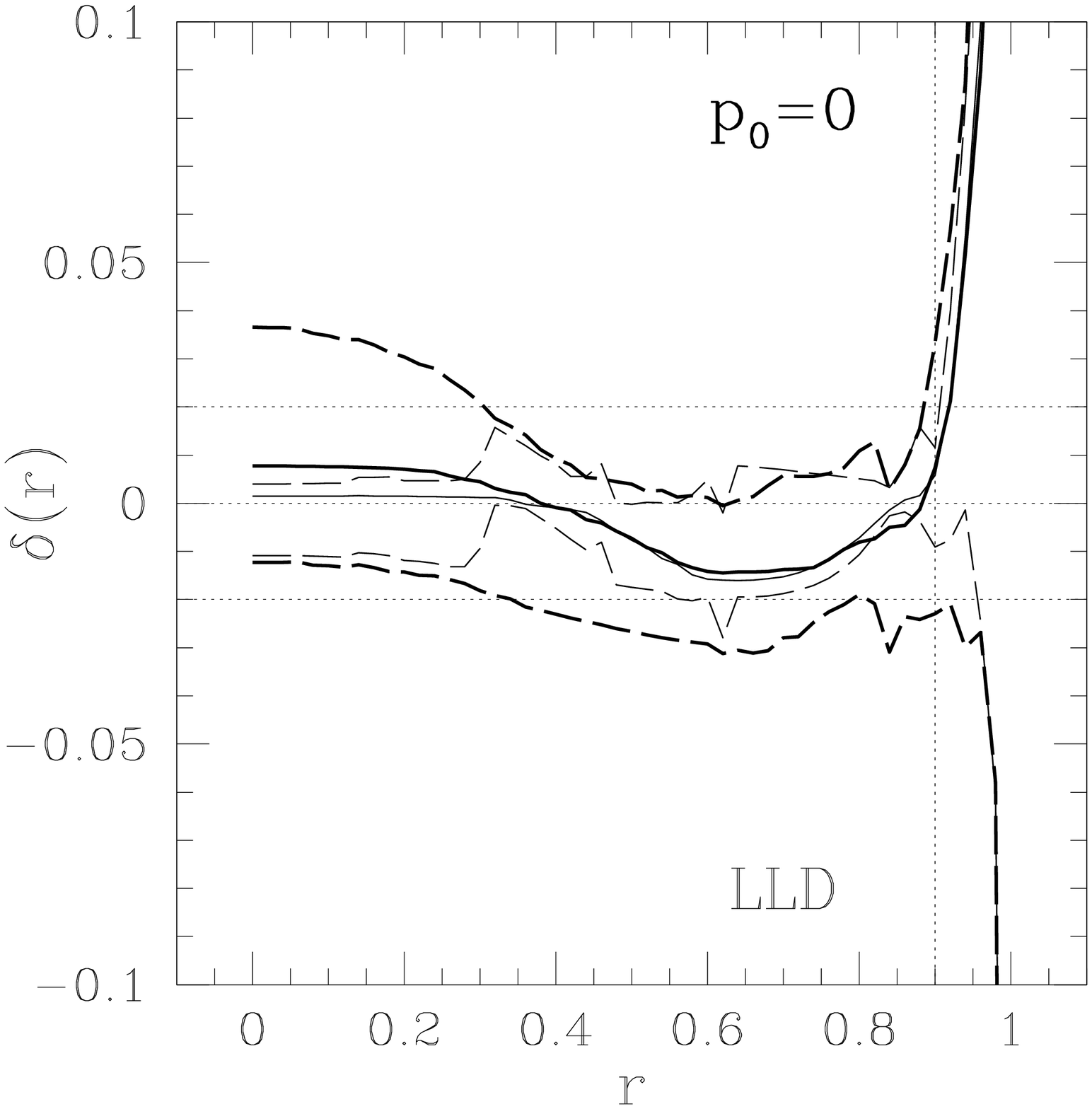}{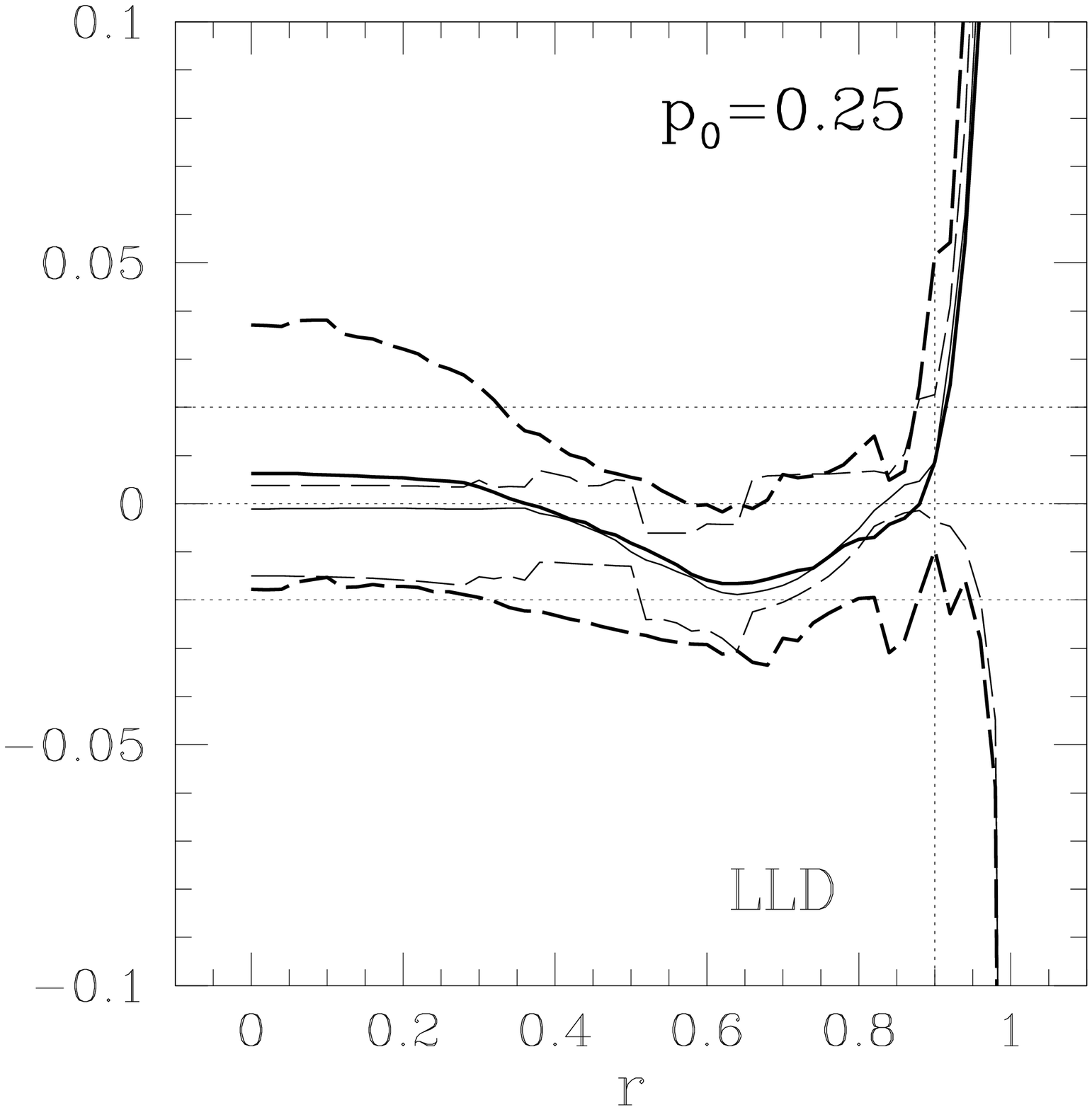}{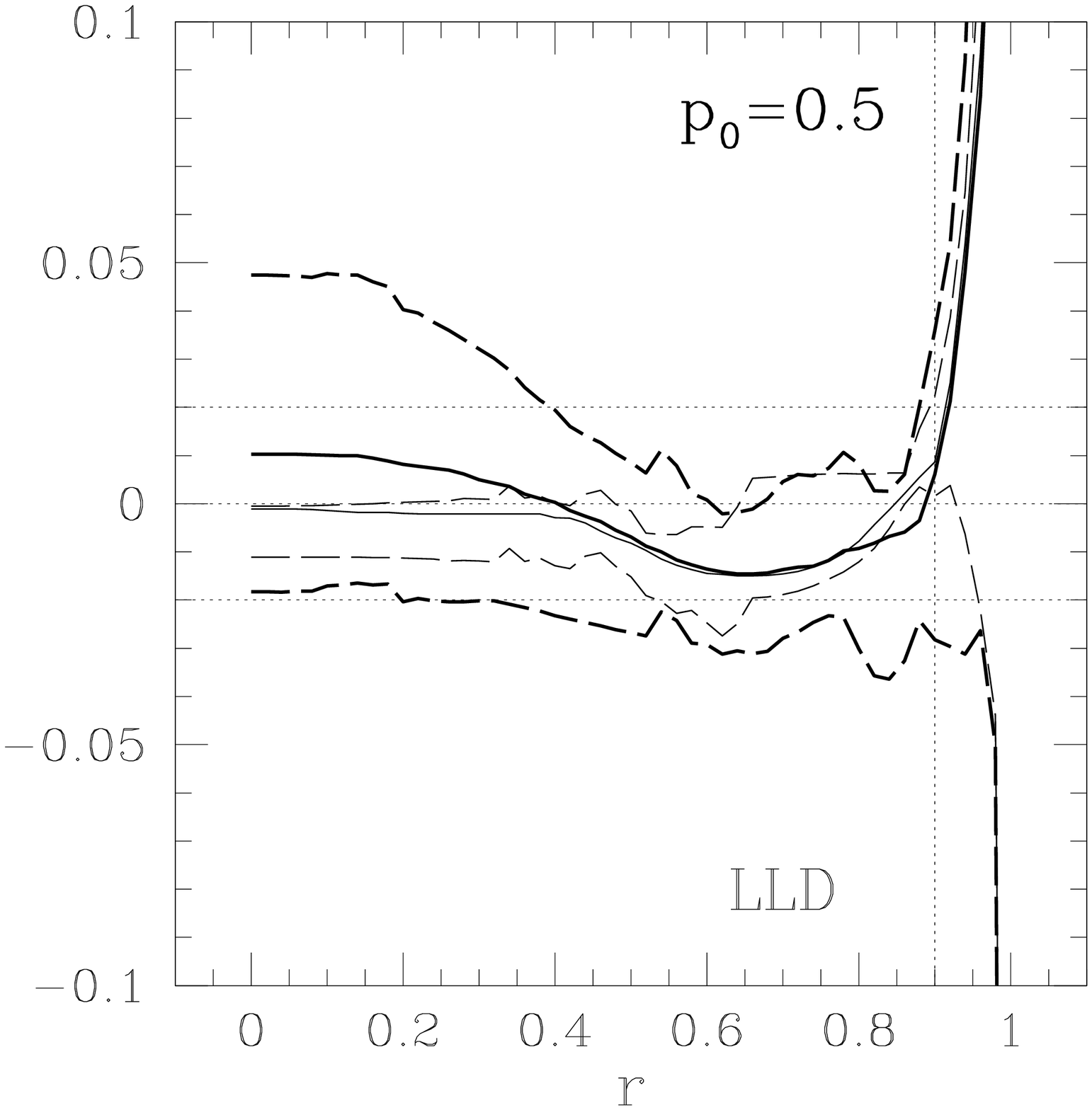}{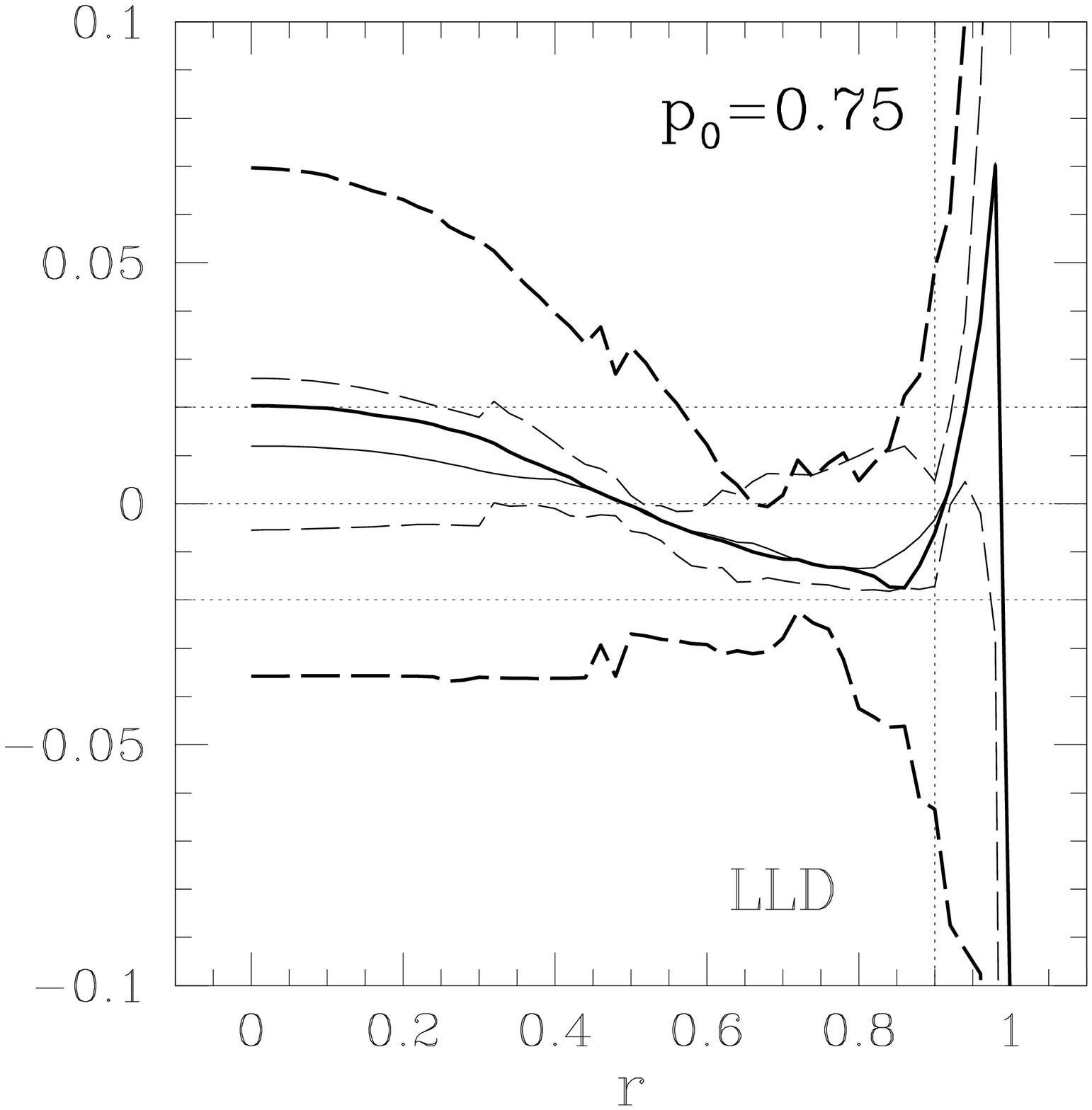}
\caption{Brightness profile inversion accuracy from center
to limb of lensed star, for $\epsilon_0=10\,$. Bold solid
line: bias $\delta_B\,$; thin solid line:``median median"
$\delta_{MM}\,$; bold dashed lines: total errorbars; thin
dashed lines: systematic errorbars. Dotted lines:
$\delta=0,\,\pm2\%$ and $r=0.9$. Multiply total errorbars
by inverted profile to obtain absolute errorbars. Columns
correspond to marked impact parameters $p_0$. Upper panels:
PCA inversion; lower panels: LLD inversion.}
\label{fig:deve10}
\enf

\bef[t]
\epsscale{0.65}
\plotone{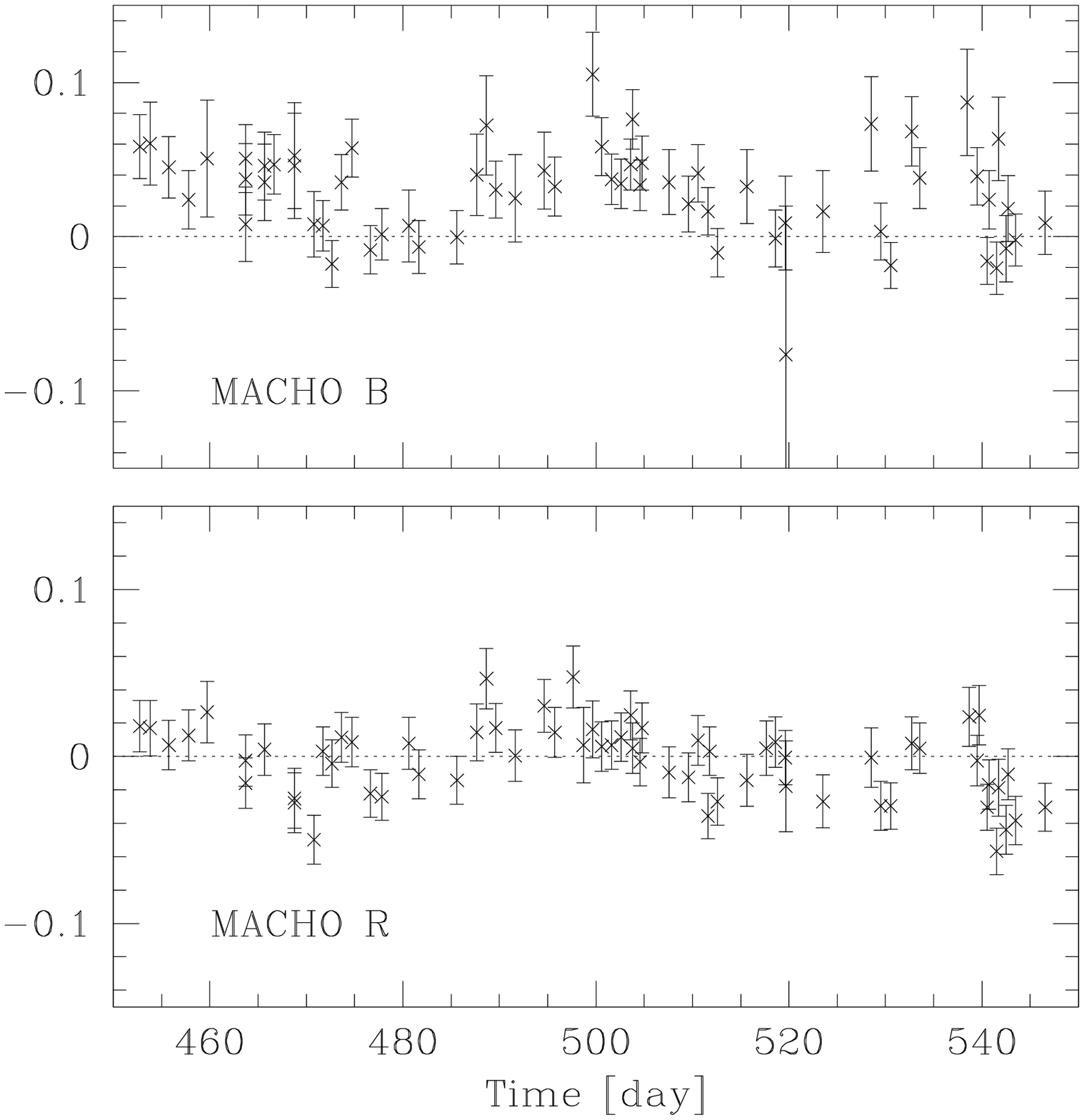}
\caption{Baseline detail of the MACHO B (upper panel) and MACHO R (lower
panel) light curves of the MACHO Alert 95-30 event. Vertical axes
correspond to relative flux deviations from the best-fit microlensing
solution baselines (marked by dotted lines). Note the correlated
variability as well as the systematically positive MACHO B deviation.
Following \citet{alc97d}, time is measured from JD 2,448,623.50 .}
\label{fig:machobase}
\epsscale{1}
\enf

\bef[t]
\plotone{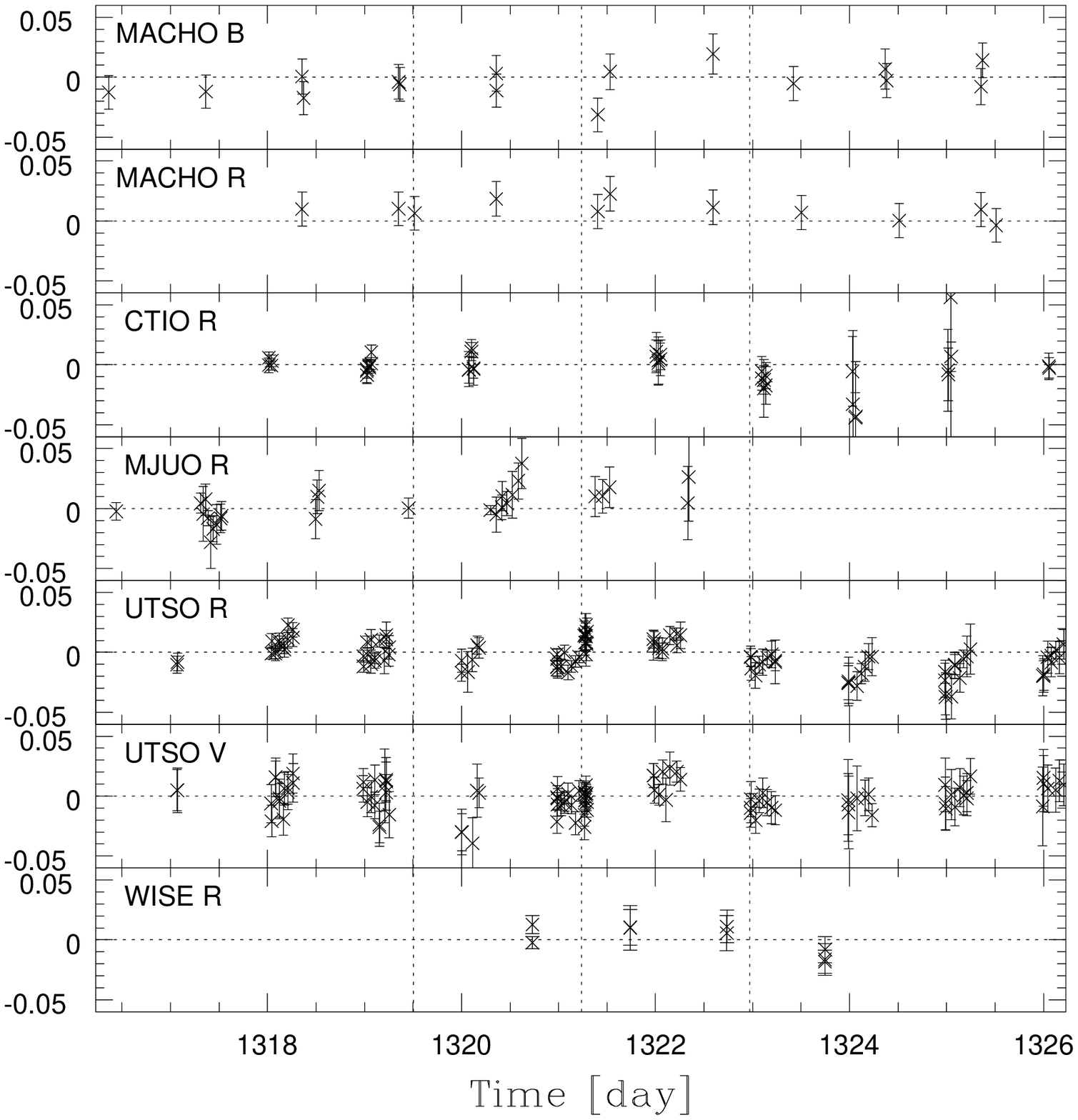}
\caption{Relative residuals $F_i/F(l_i)-1$ of the best-fit microlensing
solution close to the peak of the MACHO Alert 95-30 event. Individual
panels correspond to light curves marked in the left corners. The time of
closest approach $t_0$ and limb crossing times of the best-fit solution
are marked by vertical dotted lines (time is measured from JD
2,448,623.50) .}
\label{fig:residpeak}
\epsscale{1}
\enf

\bef[t]
\epsscale{0.7}
\plotone{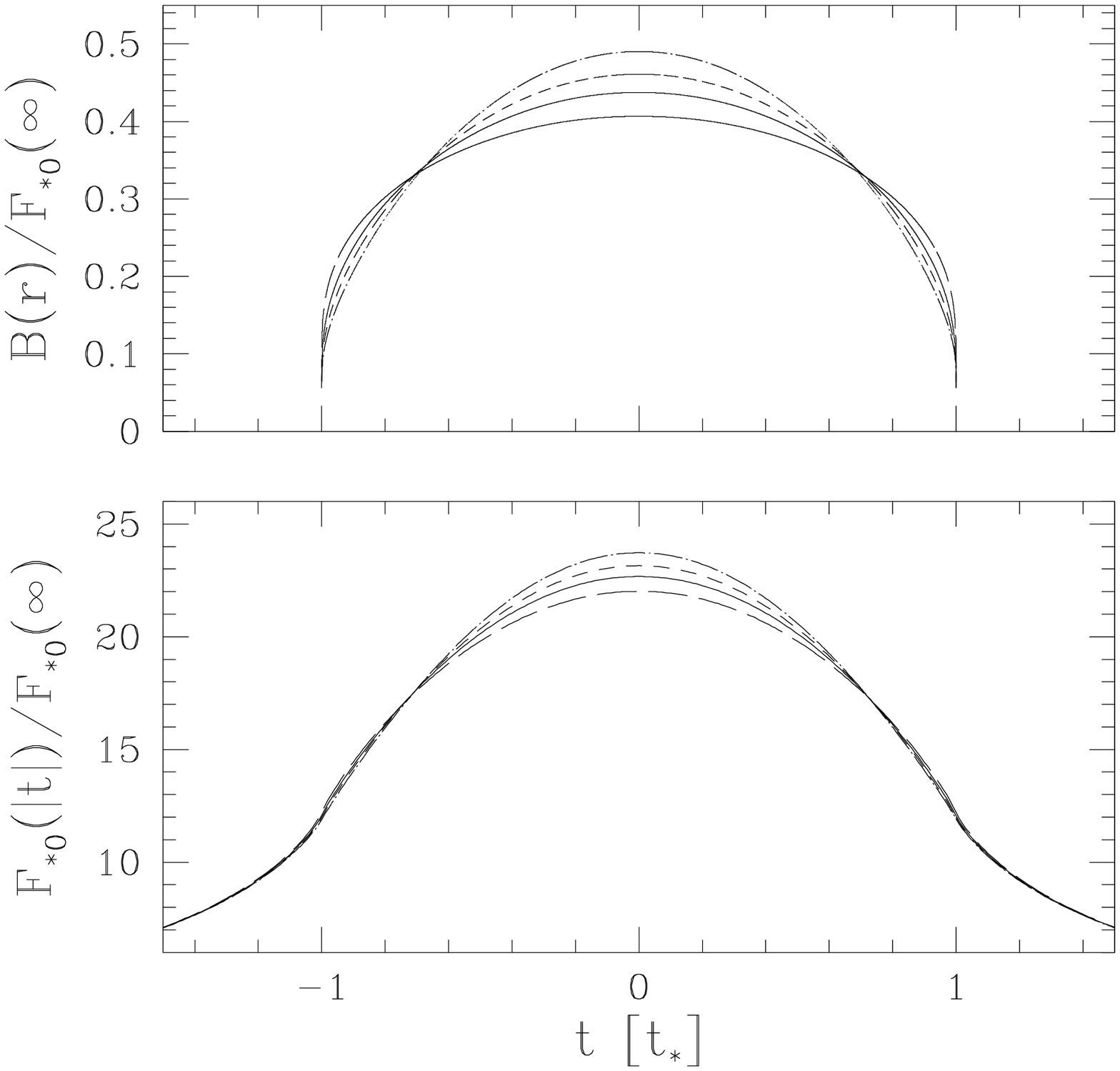}
\caption{Lower panel: $BV\!RI$ amplification light curves (from top to
bottom at $t=0$) of a microlensing transit of a T=3750 K, $\log{g}=0.5$
red giant by an $\epsilon=10$ lens with a zero impact parameter. Upper
panel: corresponding brightness profiles normalized to unit flux (same
sequence at $t=0$), horizontal scale matches the lens position.}
\label{fig:lccomp}
\epsscale{1}
\enf

\bef[t]
\epsscale{0.55}
\plotone{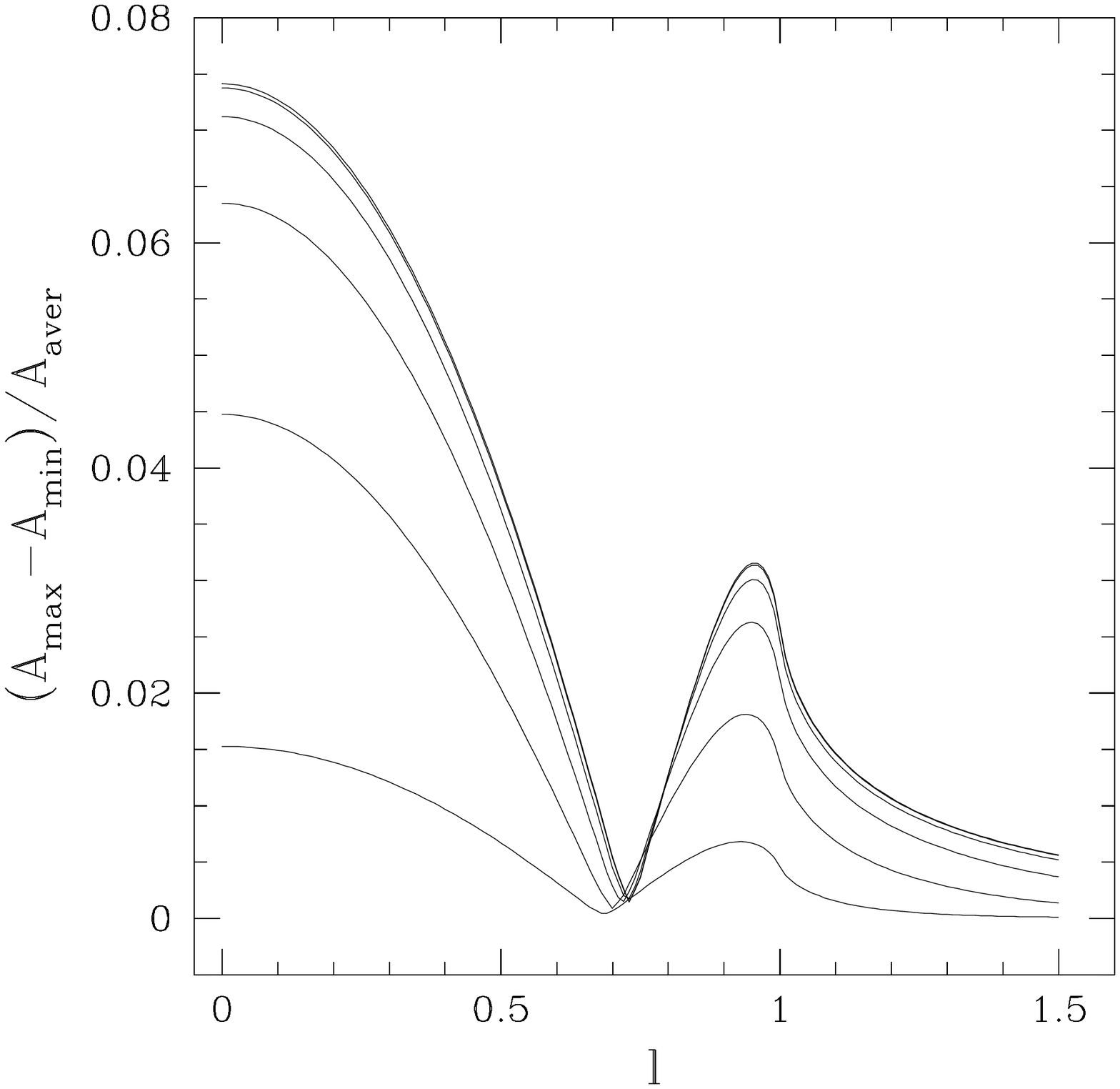}
\caption{Broadband microlensing chromaticity: relative variation of
microlensing amplification from Figure~\ref{fig:lccomp} with filter
passband, as a function of lens position (in source radius units).
Curves correspond to Einstein radii $\epsilon=0.2,0.5,1,2,5,10\,$
(from lowest curve).}
\label{fig:lcchrom}
\epsscale{1}
\enf

\bef[t]
\plottwo{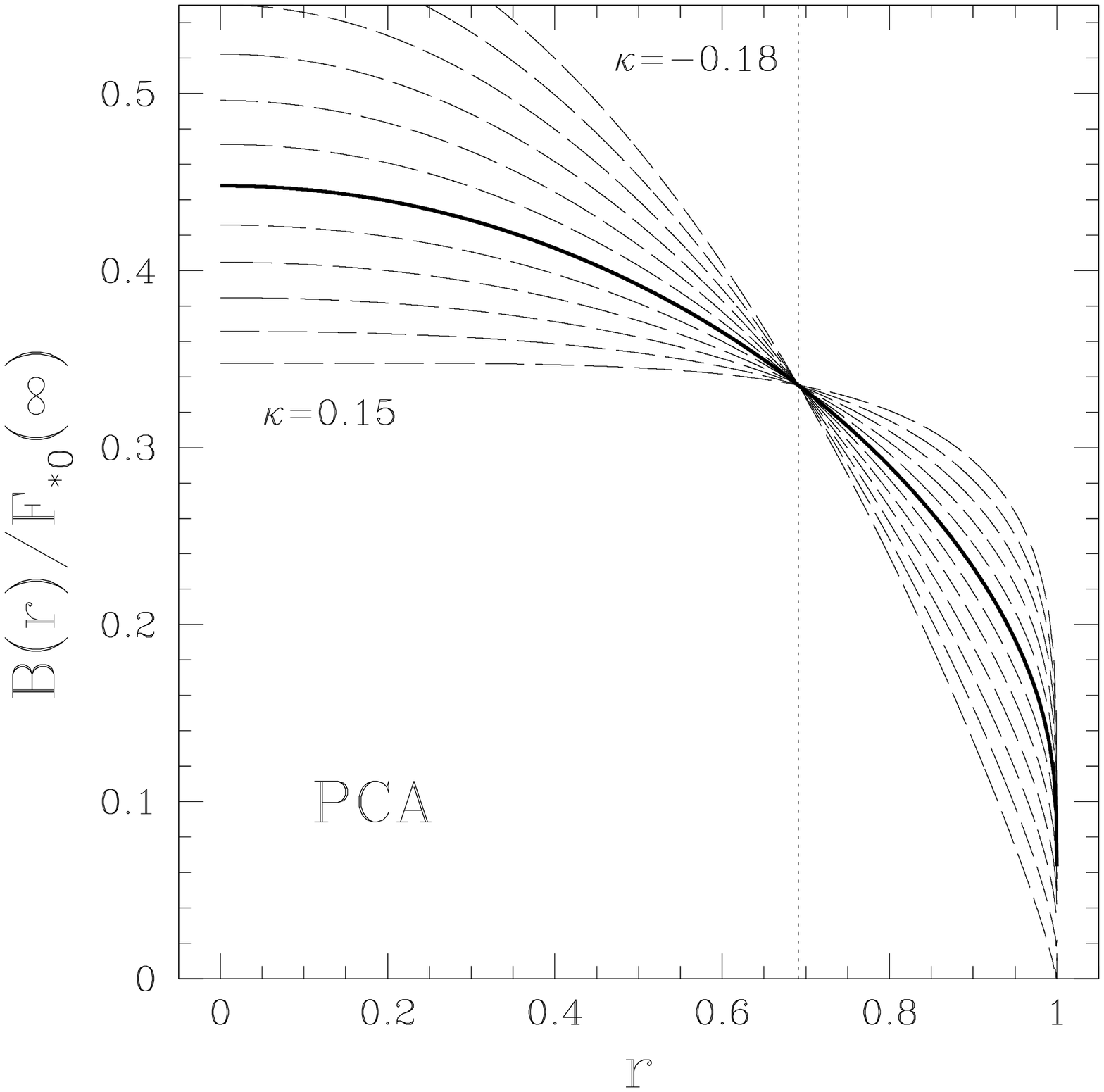}{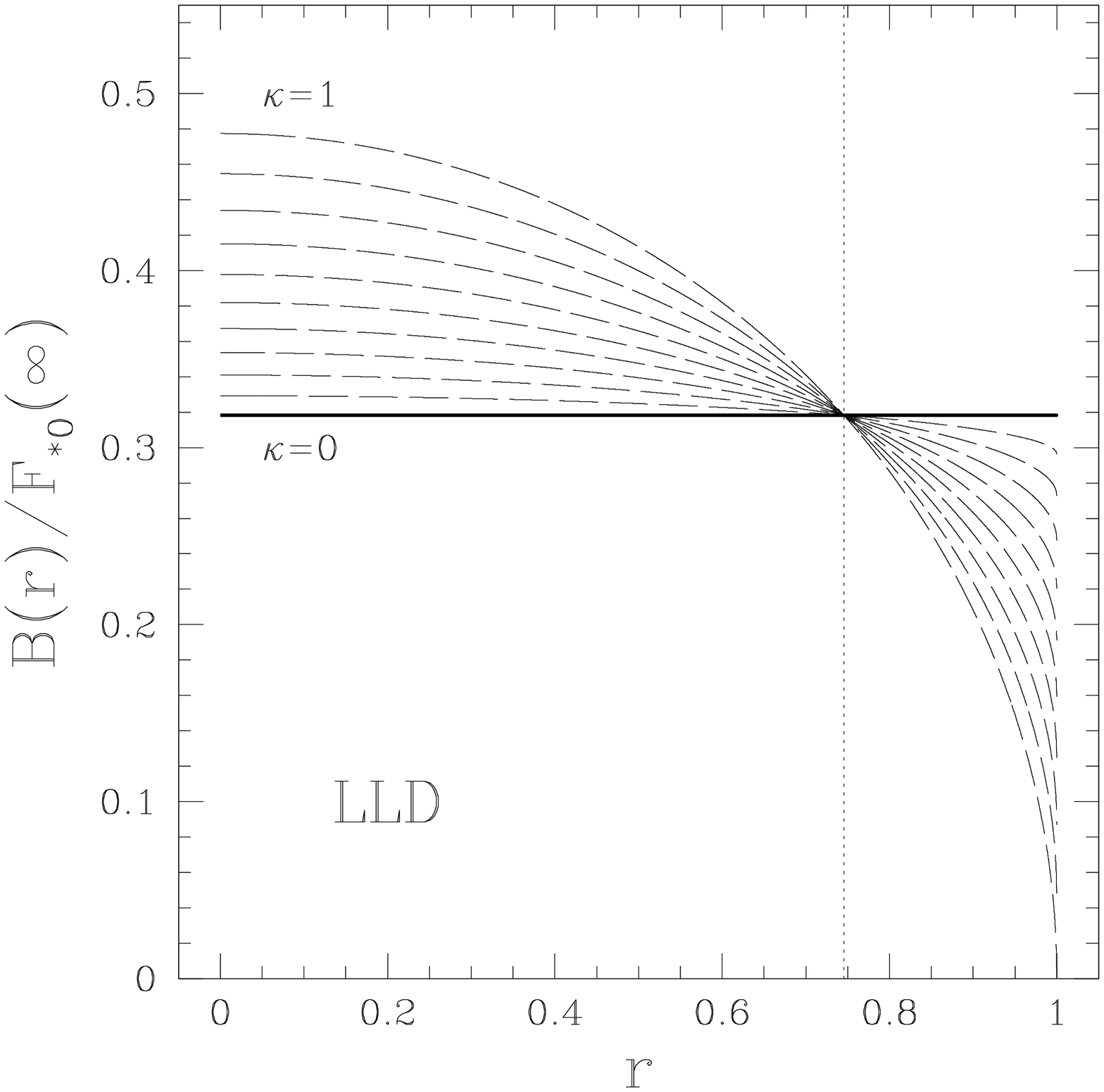}
\caption{Full parameter range of limb darkening profiles using the
2-term PCA (left panel) and LLD bases (right panel), normalized to
unit flux and plotted on same scale as Figure~\ref{fig:profiles}.
Limb darkening parameter values $\kappa$ are spaced by 0.03 in the
PCA model, by 0.1 in the LLD model. Solid curves correspond to
$\kappa=0$ in each model (first basis term). The vertical dotted
line marks the common intersection point, characteristic of each
model.}
\label{fig:pcalldprof}
\enf

\bef[t]
\epsscale{0.7}
\plotone{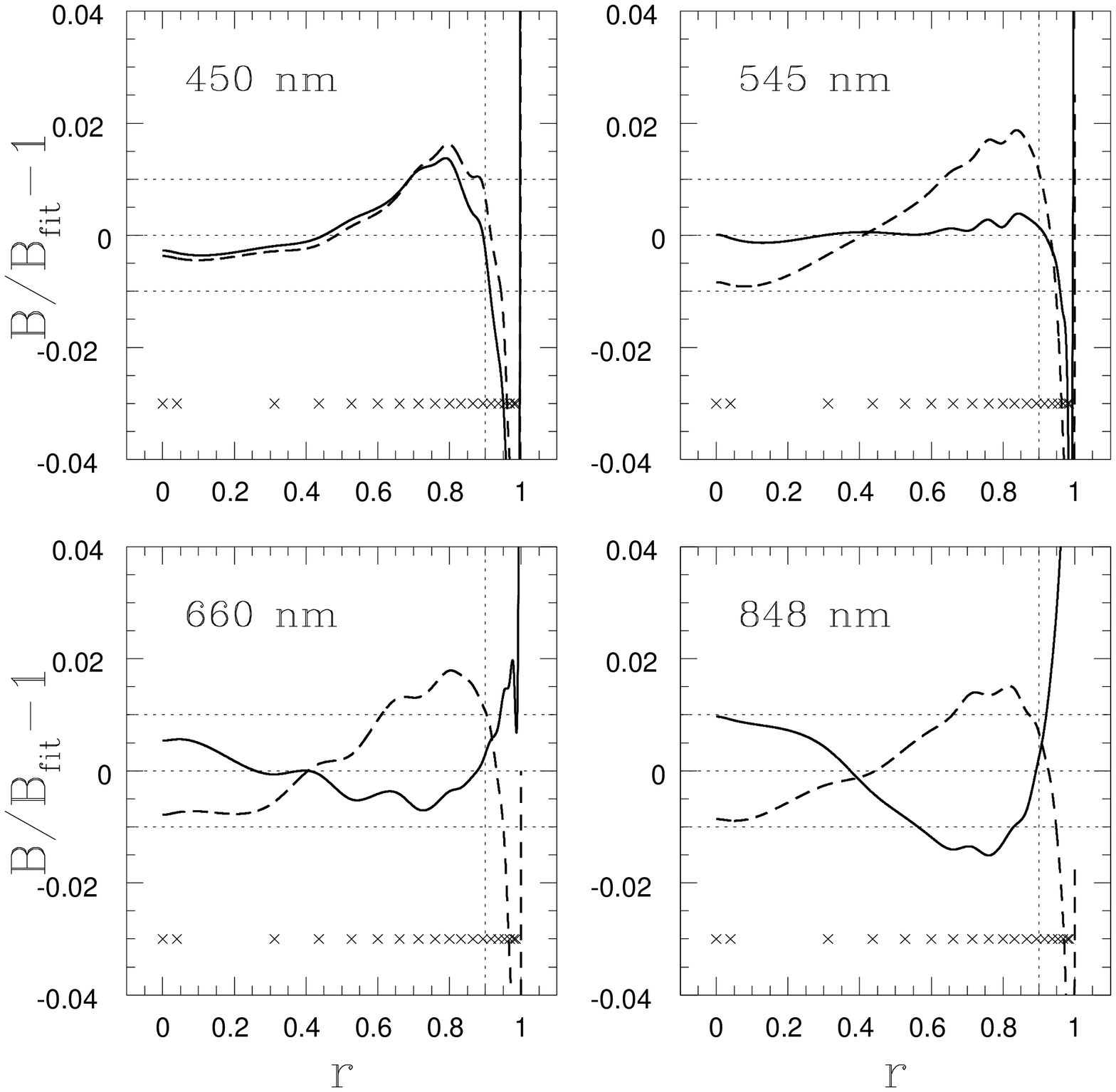}
\caption{Comparison of PCA (solid line) and LLD (dashed line) residuals
of fits to measured solar limb darkening data \citep{mitchell59} at four
selected wavelengths as denoted. Radial positions of the data points are
marked at the lower edge of the plots. Dotted lines serve for
orientation: $B/B_{fit}-1=0,\,\pm1\%$ and $r=0.9\,$.}
\label{fig:solarplot}
\epsscale{1}
\enf

\clearpage
\begin{table}
\vspace{-0.1in}
\scriptsize
\begin{center}

\begin{tabular}{ccccccc}
\tableline
\tableline
Profile & \multicolumn{3}{c}{PCA fit} & \multicolumn{3}{c}{LLD fit}\\
& Parameter & Deviation & Excess flux & Parameter & Deviation & Excess
flux\\
& [$10^{-2}$] & [$10^{-2}$] & [$10^{-4}$] & & [$10^{-2}$] & [$10^{-4}$]\\
\tableline
B3500g00  &  -4.46  &  0.68  &   0.61  &  0.970  &  2.98  &   74.61 \\
B3500g05  &  -4.96  &  0.56  &   1.06  &  0.985  &  3.00  &   75.32 \\
B3500g10  &  -5.18  &  0.44  &   1.52  &  0.993  &  2.94  &   73.96 \\
B3750g00  &  -4.67  &  0.47  &  -1.84  &  0.977  &  2.85  &   69.68 \\
B3750g05  &  -5.19  &  0.33  &  -1.48  &  0.993  &  2.87  &   70.47 \\
B3750g10  &  -5.41  &  0.23  &  -1.06  &  1.000  &  2.85  &   70.00 \\
B4000g05  &  -4.51  &  0.19  &  -1.41  &  0.977  &  2.30  &   55.46 \\
B4000g10  &  -4.84  &  0.25  &  -1.09  &  0.988  &  2.29  &   55.09 \\
V3500g00  &  -2.89  &  0.12  &   0.74  &  0.931  &  1.81  &   44.15 \\
V3500g05  &  -2.99  &  0.14  &   0.97  &  0.934  &  1.83  &   44.55 \\
V3500g10  &  -2.96  &  0.14  &   1.14  &  0.933  &  1.83  &   44.83 \\
V3750g00  &  -1.75  &  0.27  &  -0.65  &  0.898  &  1.32  &   30.96 \\
V3750g05  &  -1.83  &  0.32  &  -0.31  &  0.901  &  1.28  &   30.31 \\
V3750g10  &  -1.79  &  0.33  &  -0.12  &  0.900  &  1.26  &   29.81 \\
V4000g05  &  -0.75  &  0.51  &  -0.46  &  0.871  &  0.79  &   18.21 \\
V4000g10  &  -0.77  &  0.53  &  -0.26  &  0.872  &  0.78  &   17.86 \\
R3500g00  &  -0.25  &  0.23  &   0.94  &  0.853  &  0.94  &   22.06 \\
R3500g05  &  -0.06  &  0.11  &   0.73  &  0.847  &  1.02  &   24.02 \\
R3500g10  &   0.21  &  0.11  &   0.53  &  0.837  &  1.07  &   25.36 \\
R3750g00  &   1.32  &  0.46  &   0.91  &  0.808  &  0.31  &    5.41 \\
R3750g05  &   1.25  &  0.42  &   0.72  &  0.810  &  0.34  &    6.59 \\
R3750g10  &   1.21  &  0.35  &   0.38  &  0.810  &  0.39  &    8.30 \\
R4000g05  &   2.68  &  0.35  &   0.37  &  0.765  &  0.26  &   -1.54 \\
R4000g10  &   2.68  &  0.35  &   0.32  &  0.765  &  0.26  &   -1.39 \\
I3500g00  &   4.78  &  0.16  &   0.15  &  0.694  &  0.48  &   -3.97 \\
I3500g05  &   5.04  &  0.28  &  -0.10  &  0.685  &  0.46  &   -2.78 \\
I3500g10  &   5.30  &  0.38  &  -0.27  &  0.675  &  0.46  &   -2.24 \\
I3750g00  &   5.77  &  0.09  &   0.28  &  0.663  &  0.74  &  -11.77 \\
I3750g05  &   5.79  &  0.16  &  -0.16  &  0.662  &  0.66  &  -10.40 \\
I3750g10  &   5.80  &  0.25  &  -0.65  &  0.660  &  0.58  &   -8.80 \\
I4000g05  &   6.74  &  0.33  &  -0.65  &  0.629  &  0.77  &  -12.94 \\
I4000g10  &   6.76  &  0.36  &  -0.88  &  0.628  &  0.74  &  -12.42 \\
\tableline
\normalsize
\end{tabular}
\vspace{0.1in}
\caption{List of model brightness profiles used in this work. Adopted
profile names indicate the filter, effective temperature and surface
gravity of the model atmosphere. Following columns: best-fit PCA and
LLD limb darkening parameters $\kappa$, normalized r.m.s. deviations
$\delta_{fit}$ and relative excess fluxes $\Delta F/F$ of the fits.
\label{tab:fits}}
\end{center}
\end{table}

\end{document}